\documentclass[]{aa}

\usepackage{graphicx}
\usepackage{enumitem}
\usepackage{color}
\usepackage{txfonts}
\usepackage{bbm}
\newcommand{\kms}{{ km~s$^{-1}$}}
\newcommand{\ramses}{{\sc Ramses}}

\begin{document} 

\title{Distortions of the Hubble diagram: Line-of-sight signatures of local galaxy clusters }
 
   \author{Jenny G. Sorce\inst{1,2,3}\fnmsep\thanks{jenny.sorce@univ-lille.fr}
          \and
          Roya Mohayaee\inst{4,5}
          \and
          Nabila Aghanim\inst{2}
          \and
          Klaus Dolag\inst{6,7}
          \and
          Nicola Malavasi\inst{8}
          }
          
   \institute{Univ. Lille, CNRS, Centrale Lille, UMR 9189 CRIStAL, F-59000 Lille, France
         \and
            Universit\'e Paris-Saclay, CNRS, Institut d'Astrophysique Spatiale, 91405, Orsay, France
         \and
             Leibniz-Institut f\"{u}r Astrophysik, An der Sternwarte 16, 14482 Potsdam, Germany
           \and
             CNRS, UPMC, Institut d'Astrophysique de Paris, 98 bis Bld Arago, Paris, France
             \and
             Rudolf Peierls Centre for Theoretical Physics, University of Oxford, Parks Road, Oxford OX1 3PU, United Kingdom
             \and
             University Observatory Munich, Scheinerstr. 1, 81679 M\"{u}nchen, Germany
             \and
              Max-Planck Institut f\"{u}r Astrophysik, Karl-Schwarzschild Str. 1, D-85741 Garching, Germany
             \and
             Max Planck Institute for Extraterrestrial Physics, Giessenbachstrasse 1, 85748 Garching, Germany}
             
   \date{Received XX XX, 2023; accepted XX XX, XXXX}

 
  \abstract
   {The Universe expansion rate is modulated around local inhomogeneities due to their gravitational potential. Velocity waves are then observed around galaxy clusters in the Hubble diagram. 
This paper studies them in a $\sim$738~Mpc wide, with 2048$^3$ particles, cosmological simulation of our cosmic environment (a.k.a. CLONE: Constrained LOcal \& Nesting Environment Simulation). For the first time, the simulation shows that velocity waves that arise in the lines-of-sight of the most massive dark matter halos agree with those observed in local galaxy velocity catalogs in the lines-of-sight of Coma and several other local (Abell) clusters. For the best-constrained clusters such as Virgo and Centaurus, i.e. those closest to us, secondary waves caused by galaxy groups, further into the non-linear regime, also stand out. This match is not utterly expected given that before being evolved into a fully non-linear z=0 state, assuming $\Lambda$CDM, CLONE initial conditions are constrained with solely linear theory, power spectrum and highly uncertain and sparse local peculiar velocities. Additionally, Gaussian fits to velocity wave envelopes show that wave properties are tightly tangled with cluster masses. This link is complex though and involves the environment and formation history of the clusters. A proposed metric, measuring the distance between the observed and {several} re-centered simulated lines-of-sight, waves included, is shown to be capable of providing a tight mass range estimate for massive local clusters.
Using machine learning techniques to grasp more thoroughly the complex wave-mass relation, velocity waves could in the near future be used to provide additional and independent mass estimates from galaxy dynamics within large cluster radii.}
 

\maketitle


\section{Introduction} 

As the largest gravitationally bound structures in the Universe, galaxy clusters bear imprints of the cosmic growth visible through the distribution and motion of galaxies in their surrounding environment \citep[see][for a review and references therein]{2012ARA&A..50..353K}. They constitute therefore powerful complementary probes to supernovae and baryon acoustic oscillations in testing theories explaining cosmic acceleration origin \citep[see][for a review]{2013PhR...530...87W}. Relations between halo masses and observables (optical galaxy richness, Sunyaev-Zel'dovich effect, X-ray luminosity) must however be calibrated beforehand to study the evolution of the cluster mass function. Our capacity to discriminate among cosmological models is thus tightly linked to the accuracy of cluster mass estimates. However, most of the cluster matter content is not directly visible making their mass estimates a particularly challenging task \citep[see for a review][]{2019SSRv..215...25P,2016A&A...594A..24P}. \\

 With future imaging surveys to come (\citealp[LSST,][]{2006APS..APR.I7004B}; \citealp[Euclid,][]{2008dde..confE..33P}; \citealp[WFIRST,][]{2012arXiv1208.4012G}), stacked weak lensing measurements will certainly provide the best cluster mass estimates, i.e. with the 1\% accuracy required  \citep{2006MNRAS.372..758M} but limited to small radii around clusters. Independent virial mass estimators \citep{1985ApJ...298....8H}, hydrostatic estimators for galaxy population \citep{1997ApJ...485L..13C} or velocity caustics \citep[boundaries between galaxies bound to and escaping from the cluster potential,][]{1999MNRAS.309..610D} constitute complementary tools once calibrated. Their calibration suffers though from the influence of baryonic physics and galaxy bias on velocity fields and dispersion profiles. Perhaps velocity caustics are less prone to such systematics \citep{1999MNRAS.309..610D} explaining their recent increased popularity. Galaxy clusters can indeed be seen as disrupters of the expansion, thus creating a velocity wave first mentioned by \citet{1981ApJ...246..680T} as a triple-value region\footnote{Such an appellation derives directly from the fact that in a distorted Hubble diagram, galaxies at three distinct distances, $d$, share a similar velocity value whereas in an unperturbed diagram, these galaxy velocities would differ precisely because of the expansion proportional to H$_0~\times~d$.} whose properties (mostly height and width) depend on the cluster mass. Combined with infall models \citep{1984ApJ...281...31T,2005ApJ...635L.113M}, velocities of galaxies in the infall zones constitute thus good mass proxies for galaxy clusters shown to be in good agreement with virial mass estimates \citep{2015AJ....149...54T}. They have been used in different studies to retrieve the total amount of dark matter in groups and clusters as well as to detect groups \citep[e.g.][]{2013MNRAS.429.2264K,2013MNRAS.429.2677K}. They also permitted the finding of interlopers in galaxy clusters and of wrongly assigned distances to galaxies \citep{2004A&A...418..393S,2010A&A...520A..30M}. Moreover, \citet{2014MNRAS.445.1885Z} showed that the wave shape is a nice complementary probe: for instance, f(R) modified gravity models enhance the wave height (infall velocity) and broaden its width (velocity dispersion). This translates into a higher mass when considering a $\Lambda$CDM framework. Subsequently, it would lead to cosmological tensions between $S_8$ values measured with the cosmic microwave background and with the galaxy cluster counts. Furthermore, velocity waves probe a cluster mass within radii larger than those reached with weak lensing. Subsequently, combined together, stacked weak lensing and velocity wave mass measurements hold tighter constraints on dark energy than each of them separately. Indeed, velocity waves are signatures of a tug of war between gravity and dark energy. Differences between these two independent mass estimates, one dynamic and one static, permit measuring the gravitational slip between the Newtonian and curvature potentials. This constitutes a nice test of gravity.  \\

 Given future galaxy redshift and large spectroscopic follow-up surveys (\citealp[with Euclid, ][]{2008dde..confE..33P}; \citealp[4MOST,][]{2012SPIE.8446E..0TD}; \citealp[MOONS,][]{2014SPIE.9147E..0NC}) of imaging ones, studying galaxy infall kinematics to derive better cluster dynamic mass estimates is surely the next priority. Cosmological simulations constitute critical tools to test,  understand and eventually calibrate this mass estimate method applied to galaxy cluster observations. Ideally these simulations must be constrained simulations\footnote{The initial conditions of such simulations stem from observational constraints applied to the density and velocity fields.} to properly set the zero point of the method. Namely, simulations must be designed to ensure that the simulated and observed waves match in every aspect but if the theoretical model somewhere fails and not because of, for instance, different formation histories and/or environments. We are now able to produce such simulations valid down to the cluster scale including the formation history of the clusters \citep[e.g.][]{2016MNRAS.460.2015S,2019MNRAS.486.3951S,2021MNRAS.504.2998S,2018MNRAS.478.5199S}. These simulations are thus faithful reproduction of our local environment including its clusters such as Virgo, Coma, Centaurus, Perseus and several Abell clusters. Unlike previous analytical work using clustercentric frames to compare simulated and observed velocity waves of similar mass clusters \citep{2006NewA...11..325P}, these simulations allow us to directly compare simulated and observed full lines-of-sight (as seen from the Milky-Way) that include the clusters.\\

This paper thus starts with the first comparison between line-of-sight velocity waves due to several observed local clusters and their counterparts from a Constrained LOcal \& Nesting Environment simulation (CLONE)  built within a $\Lambda$CDM framework. First, we present the numerical CLONE used in this study. Next, we compare the observed and simulated lines-of-sight that host velocity waves. To facilitate the comparisons, the background expansion is subtracted. Wave envelopes are, then, fitted to study relations between wave properties and cluster masses in a $\Lambda$CDM cosmology. Before concluding, a metric, measuring the distance between the observed and {several} re-centered simulated lines-of-sight, waves included, is proposed in order to deduce a mass range estimate of the clusters they host.


\section{The CLONE simulation}

 Constrained simulations are designed to match the local large-scale structure around the Local Group. Several techniques have been developed to build the initial conditions of such simulations \citep[e.g.][]{2010arXiv1005.2687G,2013MNRAS.429L..84K,2013MNRAS.432..894J} with density, velocity or both constraints. Here we use the technique whose details (algorithms and steps) are described in \citet{2016MNRAS.460.2015S,2018MNRAS.478.5199S}. Local observational data used to constrain the initial conditions are distances of galaxies and groups \citep{2013AJ....146...86T,2017MNRAS.469.2859S} converted to peculiar velocities \citep{2016MNRAS.455.2078S,2018MNRAS.476.4362S} that are bias-minimized \citep{2015MNRAS.450.2644S}. We showed that constrained simulations obtained from this particular technique, a.k.a. the CLONES \citep{2021MNRAS.504.2998S}, are, to our knowledge, currently the sole replicas of the local large-scale structure that include the largest local clusters using only galaxy peculiar velocities as constraints. Namely, the cosmic variance of replicas (a constrained simulation set) is effectively reduced within a 200~Mpc radius centered on the Local Group down to the cluster scale, i.e. 3-4 Mpc, \citep{2016MNRAS.460.2015S} with respect to that of a random simulation set. Galaxy clusters (such as Virgo, Centaurus, Coma) have masses in agreement with observational estimates {\citep{2018MNRAS.478.5199S,2024arXiv240201834H}}. Several ensuing studies focused in particular on the galaxy clusters. These studies confirmed the necessity of using CLONES to get a high-fidelity Virgo-like cluster. Additionally, they confirmed observationally-based formation scenarios of the latter \citep{2018A&A...614A.102O,2019MNRAS.486.3951S,2021MNRAS.504.2998S}. They also highlighted projection effects on its hydrostatic mass, in particular due to a group on the same line-of-sight as Virgo \citep{2023arXiv231002326L}{, and dynamic state effects on its splashback radius measurement \citep{2024arXiv240318648L}}. Another work looked at the cosmic web surrounding the Coma cluster and recovered the observed filaments \citep{2023A&A...675A..76M}. One of the latest studies determined the peculiarities of the distribution of the local clusters, in terms of mass and position, with respect to that of clusters in random patches \citep{2023A&A...677A.169D}. \\
 
To actually probe a large range of velocities in the infall zones, the CLONE for the present study needs to have a sufficient resolution to simulate, with a hundred particles at z=0, halos of intermediate mass ($\sim$10$^{11}$-10$^{12}~$M$_\odot$). Moreover, \citet{2010MNRAS.407.1487P} showed that a dark matter only simulation is sufficient for velocity wave studies. They indeed found similar results when adding baryons. They found no major difference regarding cluster outskirt dynamics on these scales. The constrained initial conditions of the CLONE contain thus 2048$^3$ dark matter particles in a $\sim$738~Mpc comoving box (particle mass $\sim$10$^9$~M$_\odot$). It ran on more than 10,000 cores from z=120 to z=0 in the Planck cosmology framework \citep[$\Omega_m$=0.307 ; $\Omega_\Lambda$=0.693~;  H$_0$=67.77~\kms~Mpc$^{-1}$ and $\sigma_8$~=~0.829,][]{2014A&A...571A..16P} using the adaptive mesh refinement \ramses\ code \citep{2002A&A...385..337T}. The mesh is dynamically (de-)refined from levels 11 to 18 according to a pseudo-Lagrangian criterion, namely when the cell contains more (less) than eight dark matter particles. The initial coarse grid is thus adaptively refined up to a best-achieved spatial resolution of $\sim$2.8~kpc roughly constant in proper length (a new level is added at expansion factors $a=0.1,0.2,0.4,0.8$ up to level 18 after $a=0.8$). \\

    \begin{figure}
\flushleft
\vspace{-1.2cm}

\includegraphics[width=0.55\textwidth,trim={0.4cm 0 0 0},clip]{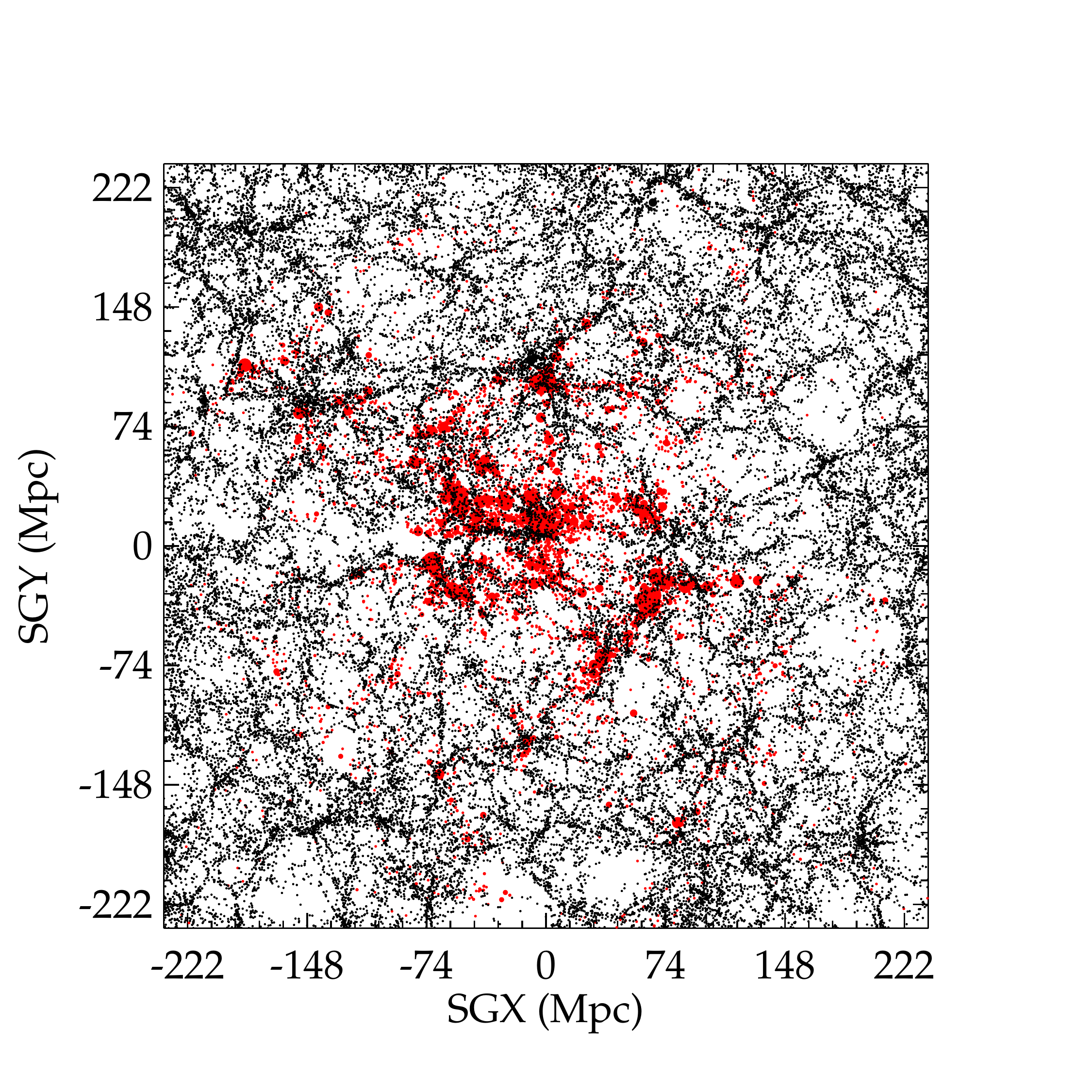}

\vspace{-0.2cm}

\caption{{$\sim$30~Mpc} thick XY supergalactic slice of the CLONE. Black dots stand for the dark matter halos (subhalos are excluded for clarity). Red dots are galaxies and groups from the grouped 2MASS Galaxy Redshift Catalog (2MRS) for comparison purposes only \citep[compressed Fingers-of-God from][]{2018A&A...618A..81T}. Indeed, only a small fraction of local galaxy observational redshifts have been used to derive peculiar velocities that were used as constraints (about $\sim$2.5\% of the 2MRS catalog). Red dot sizes are proportional to the richness of the groups. {Observed groups (large red dots) are on top of simulated dark matter halos (black dots). No large red dot appears to be overlaid on the middle of a simulated void.}}
\label{fig:global}
\end{figure}

Using the halo finder, described in \citet{2004MNRAS.352..376A} and \citet{2009A&A...506..647T}, modified to work with 2048$^3$ ($>$2$^{31}$) particles, dark matter halos and subhalos are detected in real space with the local maxima of dark matter particle density field. Their edge is defined as the point where the overdensity of dark matter mass drops below 80 times the background density. We further apply a lower threshold of a minimum of 100 dark matter particles. Fig. \ref{fig:global} shows the {$\sim$30~Mpc} thick XY supergalactic slice of the CLONE. Black (red) dots stand for the dark matter halos \citep[galaxies from the grouped 2MASS Galaxy Redshift Catalog,][]{2012ApJS..199...26H,2018A&A...618A..81T}. Red dot sizes are proportional to the richness of the groups. 2MRS galaxies are used for sole comparison purposes. 2MRS is indeed far more complete than the peculiar velocity catalog used to constrain the simulation ($\sim$2.5\% of the redshift catalog is used to derive the peculiar velocity). In fact, it shows the constraining power of the peculiar velocities that are correlated on large scales. Namely, the simulation is constrained also in regions where no peculiar velocity measurements were available and thus used as constraints. It confirms that peculiar velocity catalogs fed to our technique, to reconstruct/constrain the local density and velocity fields, do not need to be complete \citep{2017MNRAS.468.1812S}.


\section{Velocity wave}
 \begin{figure}
  \vspace{-0.cm}
\center\includegraphics[width=0.5 \textwidth]{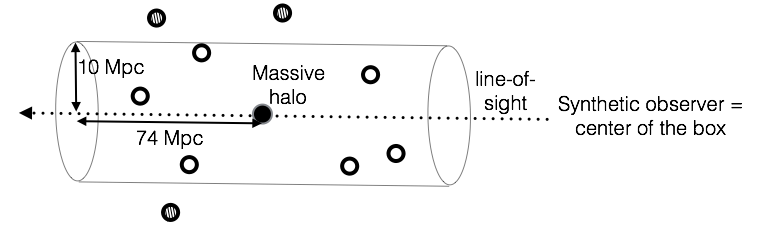}
\caption{Schema of the cylinder used to select (sub)halos whose radial peculiar velocities, derived as a function of the synthetic observer at the  simulated box center, are used to study the velocity wave arisen from the massive halo in its center. While open circles stand for selected halos, dashed circles represent excluded ones.}
\label{fig:schema}
\end{figure}

\begin{figure*}
\center
\vspace{-1.3cm}
\hspace{0.5cm}\includegraphics[width=0.5 \textwidth, trim={1cm 0 1cm 0},clip]{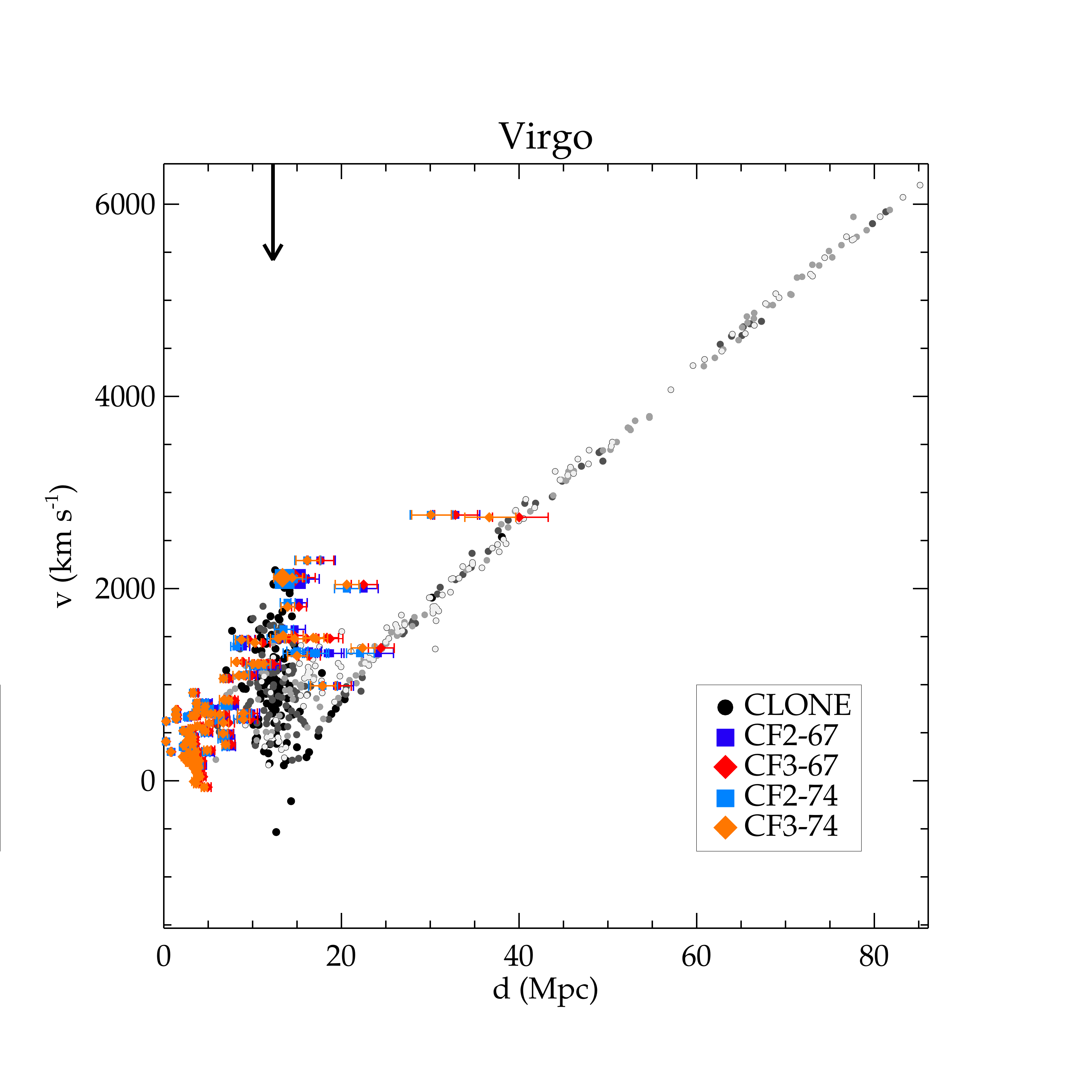}\hspace{-1cm}
\includegraphics[width=0.5 \textwidth, trim={1cm 0 1cm 0},clip]{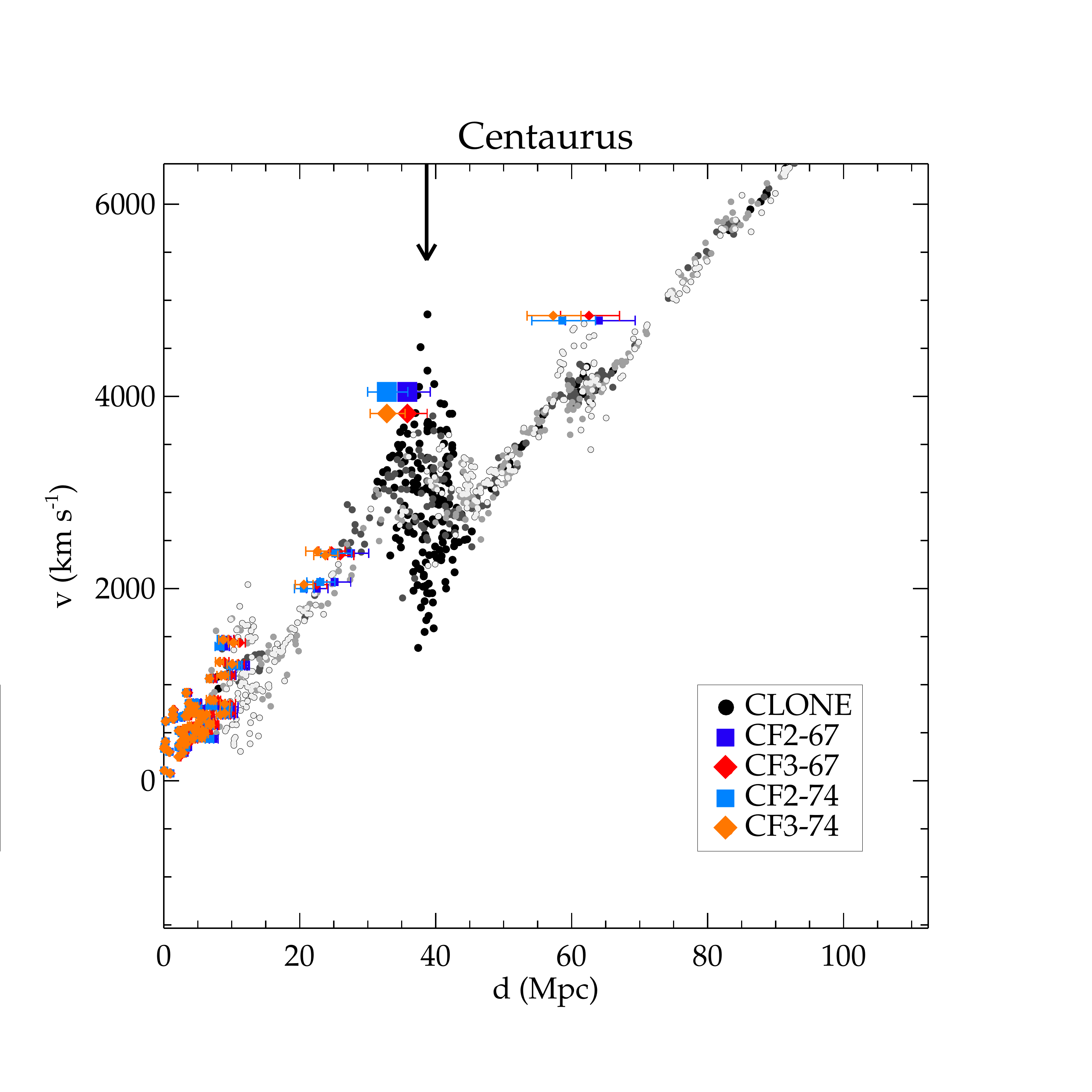}
\vspace{-1.6cm}

\hspace{0.5cm}\includegraphics[width=0.5 \textwidth, trim={1cm 0 1cm 0},clip]{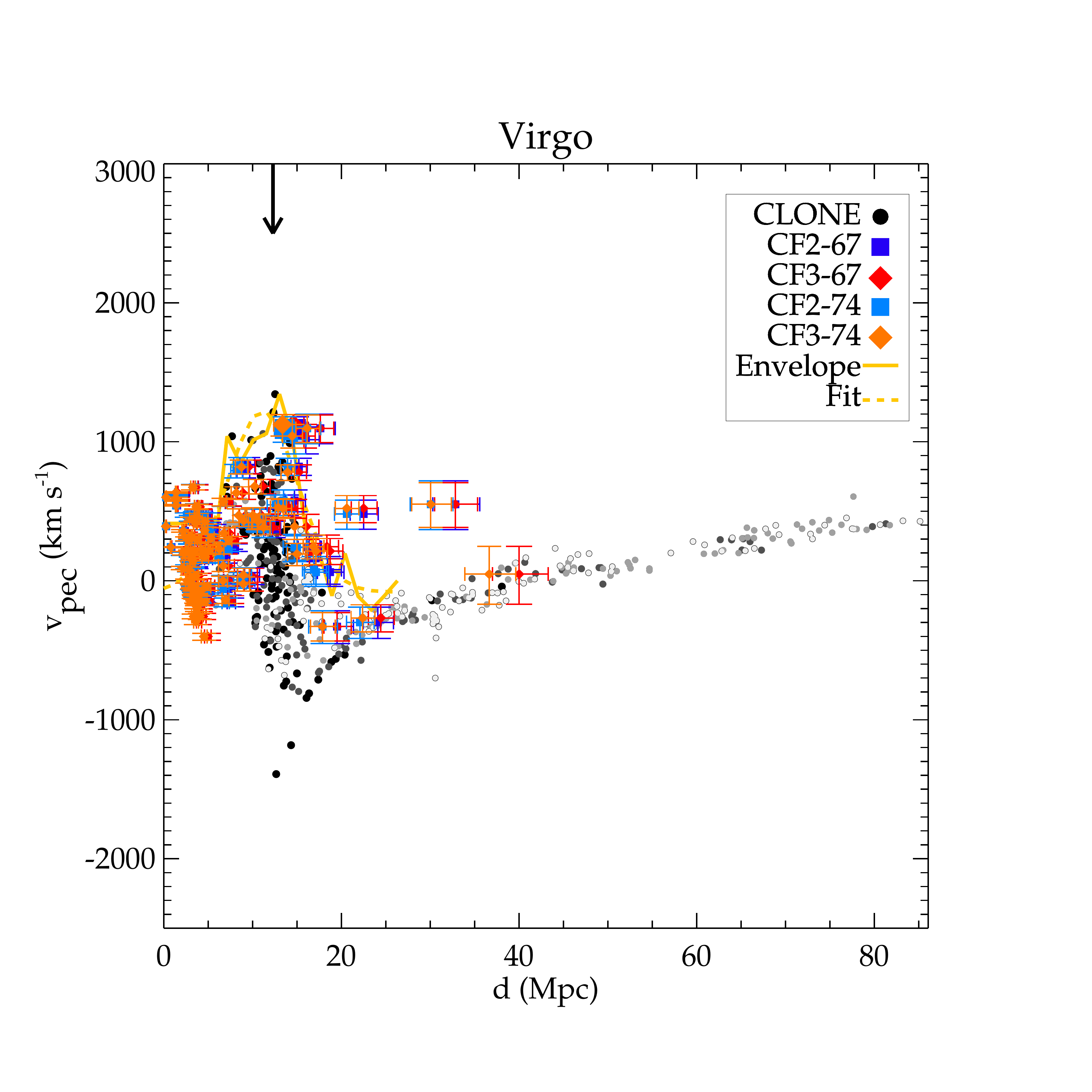}\hspace{-1cm}
\includegraphics[width=0.5 \textwidth, trim={1cm 0 1cm 0},clip]{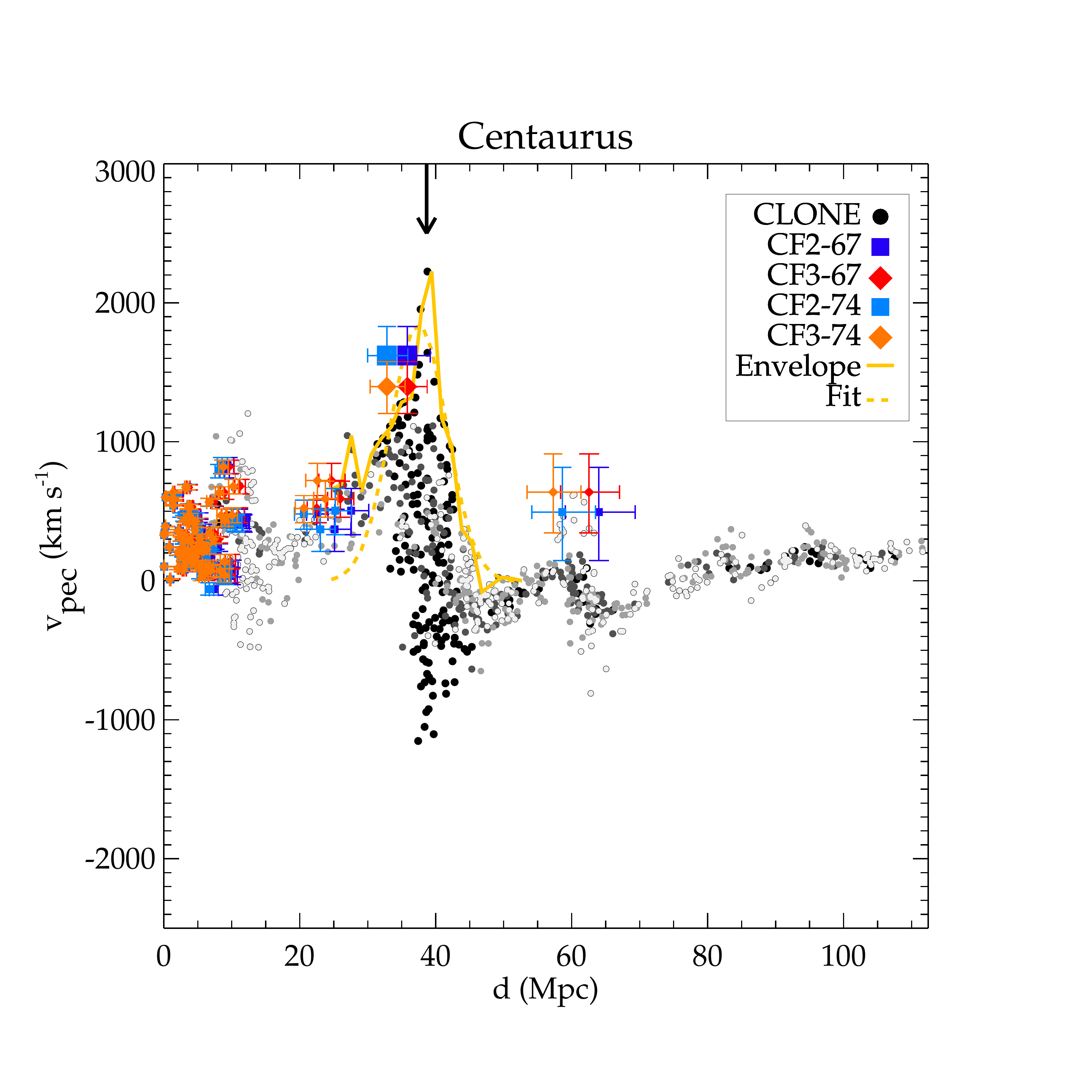}
\vspace{-0.7cm}

\caption{Radial velocities of simulated dark matter (sub)halos (black and grey) and observed galaxies (orange, blue and red) as a function of the distance from the synthetic observer and us respectively. Error bars stand for uncertainties on observational distance and velocity estimates. Orange and light blue (red and dark blue) filled squares and diamonds show observed galaxies assuming H$_0$=74 (67.77)~km~s$^{-1}$~Mpc$^{-1}$ for scaling positions. CF2 (CF3) corresponds to the second (third) catalog of the Cosmicflows project. Larger symbols are used for galaxies, with a peculiar velocity higher than 1000~\kms, identified as the closest to the simulated massive halos assuming the synthetic observer at the box center and the same Supergalactic coordinate system and orientation as the local Universe. The arrow indicates the position of the massive dark matter halo in the simulation. Names of corresponding observed clusters are given at the top of each panel. Velocity waves stand out in the different lines-of-sight and there is a nice agreement with observational datapoints for those two best-constrained clusters the closest to us. Top: Hubble diagram. Bottom: Hubble flow subtracted in the observations (H$_0$=74~km~s$^{-1}$~Mpc$^{-1}$, see the text for explanation on the zero point choice of the observational catalog) and not added in the simulation. The solid and dashed yellow lines are respectively the simulated positive-half velocity wave envelope and its Gaussian-plus-continuum fit. The color scale filling the black circles stands for their distance from the line-of-sight. From black to light grey, objects are less than 2.5, 5, 7.5 and 10~Mpc away from the line-of-sight. The dark matter halo virial masses in the simulation are M=9.8$\times$10$^{14}$M$_\odot$ and  M=9.0$\times$10$^{14}$M$_\odot$ for the Virgo and Centaurus cluster counterparts respectively.}
\label{fig:waves}
\end{figure*}

\subsection{In simulated data}

Positioning a synthetic observer at the simulation box center, we derive radial peculiar velocities for all the dark matter halos and subhalos in the z=0 catalog. We then draw lines-of-sight in the direction of each dark matter halo more massive than 5~10$^{14}$M$_\odot$. All the (sub)halos within 10~Mpc from the line-of-sight and within 74~Mpc along the line-of-sight from a given massive dark matter halo (with the center and edge of the box as upper limits) are selected to plot the latter corresponding velocity wave. Namely, as shown on Fig. \ref{fig:schema}, radial peculiar velocities, with respect to the synthetic observer, of (sub)halos within a cylinder at maximum 148~Mpc long and 20~Mpc wide are used to visualize the velocity wave caused by the massive dark matter halo in the cylinder center. Because the simulation is constrained to reproduce the local Universe, we choose not to use the periodic boundary conditions to wrap around the box edges. It will indeed not be representative of local structures. A 10~Mpc radius cylinder corresponds to about three times the virial radius of  the massive clusters under study here (M$>$5 10$^{14}$M$_\odot$). Since the goal is to study the link between velocity wave properties and cluster masses, exact masses cannot be used to define the cylinder shape. Finally, such large volumes permit probing the infall region around the massive halos. A cylinder shape is preferable to a cone shape to get an unbiased wave signal. A cone would indeed result in a distorted signal as it would probe a larger and larger region around a massive halo with the distance.

\subsection{In observational data}

Observational data are taken from the raw second and third catalogs of the Cosmicflows project \citep{2013AJ....146...86T,2016AJ....152...50T}. The second catalog containing $\sim$8000 galaxies, with a mean distance of $\sim$90~Mpc, serves as the basis to build the constraint-catalog of $\sim$5000 bias-minimized radial peculiar velocities of galaxies and groups with a mean distance of $\sim$60~Mpc. By contrast, the third catalog contains $\sim$17,000 galaxies with a mean distance of $\sim$120~Mpc. The third catalog is not used to constrained our CLONE initial conditions and thus constitute partly an independent dataset for consistency checks. More precisely, it serves the two-fold goal of extending the number of observational datapoints to be compared with the simulation and highlighting the constraining power of peculiar velocities. The latter can indeed permit the recovery of structures that are not directly probed and that are at the limit of the non-linear threshold. In the sense that there is no direct measurement in a given region but, because the latter influences the velocities of other regions (large-scale correlations), it can still be reconstructed. \\

Uncertainties on distances and radial peculiar velocities in these catalogs depend on the distance indicator used to derive the distance moduli. Error bar sizes need to be limited to see clearly velocity waves. Thus, to be able to compare with the simulated data, only galaxies with uncertainties on distance moduli smaller than 0.2~dex are retained. There remain 338 and 424 galaxies respectively from the second and third catalogs with a mean distance of $\sim$50~Mpc. These galaxies are mostly hosts of supernovae, especially those the furthest from us (distance indicator with a small uncertainty even as the distance increases). \\

To derive the radial peculiar velocities of these galaxies, we use both galaxy distance moduli ($\mu$) and observational redshifts (z$_{\mathrm{obs}}$) following \citet{2014MNRAS.442.1117D}. We add supergalactic longitude and latitude coordinates to derive galaxy Cartesian supergalactic coordinates.  A cosmological model is then required to determine peculiar velocities. While we use $\Lambda$CDM, as Cosmicflows catalog zero points are calibrated through a long process on WMAP (rather than Planck) values \citep[$\Omega_m$=0.27, $\Omega_\Lambda$=0.73, H$_0$=74~\kms~Mpc$^{-1}$,][]{2013AJ....146...86T,2016AJ....152...50T}, we have to use the same parameter values. We indeed showed that when applying the bias minimization technique to the peculiar velocity catalog of constraints, we reduce the dependence on $\Lambda$CDM cosmological parameter values \citep{2017MNRAS.469.2859S}. However, in order to be able to probe the whole velocity wave for the comparisons, we have to use the raw catalog i.e. with neither galaxy grouping nor bias minimization. Consequently, if we were to take Planck values to derive galaxy peculiar velocities, the WMAP calibration would translate into a residual Hubble flow visible in the background-expansion-subtracted Hubble diagram. Subsequently, using WMAP values for the observations:\\

\noindent\textbf{Luminosity distances, $d_{\mathrm{lum}}$}, are derived from distance modulus measurements, $\mu$, obtained via distance indicators:
 \begin{equation}
 \mu=5\mathrm{log_{10}}(d_{\mathrm{lum}}~\mathrm{(Mpc)})+25
 \end{equation}
 
 \noindent\textbf{Cosmological redshifts, z$_{\mathrm{cos}}$}, are then obtained through the equation  {\citep{1972gcpa.book.....W}}:  
   \begin{equation}
   d_{\mathrm{lum}}=(1+z_{\mathrm{cos}})\int_0^{z_{\mathrm{cos}}}\frac{cdz}{H_0\sqrt{(1+z)^3\Omega_m+\Omega_\Lambda}}
   \end{equation}
   
\noindent\textbf{Galaxy radial peculiar velocities, $v_{\mathrm{pec}}$}, are finally estimated, using the observational  z$_{\mathrm{obs}}$ and cosmological z$_{\mathrm{cos}}$ redshifts  with the following formula \citep{1972gcpa.book.....W}:
   \begin{equation}
   v_{\mathrm{pec}} = c \frac{z_{\mathrm{obs}}-z_{\mathrm{cos}}}{1+z_{\mathrm{cos}}}
   \end{equation}
 where $v_{\mathrm{pec}}$ will always refer to the \emph{radial} peculiar velocity in this paper and $c$ is the speed of light.

\subsection{Simulated versus observed data}

Assuming the synthetic observer at the box center and the simulated volume oriented similarly to the local volume, observed and simulated positions and lines-of-sight can be matched. We can only compare velocity waves born from local galaxy clusters for which infalling galaxy peculiar velocities, with uncertainties on corresponding distance moduli smaller than 0.2~dex, are available in the observed cluster surroundings. We thus select these clusters. For each simulated massive dark matter halo, the quickest way is then to search for the closest observed galaxy, in our selected above samples, with a radial peculiar velocity greater than 1000~km~s$^{-1}$ ($\sim$2$\sigma$ above the average). This is indeed a signature that it has most probably an observed cluster with a mass of at least a few 10$^{14}$M$_\odot$ as a neighbor. Whenever a simulated massive dark matter halo is within the 2$\sigma$ uncertainty of the observed galaxy distance, we select all the observed galaxies in the cylinder corresponding to the line-of-sight. For every case, there is indeed a massive observed cluster in the vicinity of the galaxies. More to the point, given the Supergalactic coordinates of the observed clusters and those of the simulated ones in the box, they match. \\

Fig. \ref{fig:waves} superimposes observed and simulated lines-of-sight that include velocity waves born from the two closest most massive local clusters. Observational data are of sufficient quality in the infall regions to warrant adequate comparisons. From left to right, galaxy clusters (dark matter halos) are at increasing distance from us (the synthetic observer). The name of the clusters is indicated at the top of each panel. Filled black and grey circles stand for simulated (sub)halos while filled light blue and orange squares and diamonds represent observed galaxies. Because the simulation was run with H$_0$~=~67.77~\kms~Mpc$^{-1}$, filled dark blue and red squares and diamonds are observed galaxies at positions rescaled with this latter value. Position differences are within about the 1$\sigma$ uncertainty on the distance. Arrows indicate the position of the most massive halos in the lines-of-sight of interest. 

In the top panels, the Hubble diagrams are distorted by the presence of massive halos. Their corresponding velocity wave or triple-value region signatures show up. The bottom panels, with the Hubble flow subtracted, equally confirm the waves. The simulated velocity waves of the two massive dark matter halos emerge in the peculiar velocity of (sub)halos plotted as a function of their distance from the synthetic observer diagrams. There is a qualitative match with the observational data points. All the more since only sparse peculiar velocities of today field galaxies and groups are used to constrain the linear initial density and velocity fields, at the positions of the latter progenitors, using solely linear theory and a power spectrum assuming a given cosmology. Then the full non-linear theory is used to evolve these initial conditions from the initial redshift down to z=0 within a $\Lambda$CDM framework. 

The signatures of Virgo West and the group around NGC4709, that are respectively beyond Virgo and Centaurus in the lines-of-sight, can also be identified as secondary waves. These smaller waves follow the highest ones representing the main clusters in both the observational and the simulated lines-of-sight. There exist a visual agreement between the observed lines-of-sight dynamical state of Virgo and Centaurus and those reproduced by CLONE. Comparisons with another constrained simulation, called SIBELIUS \citep{2022MNRAS.512.5823M}, shown in the appendix, reveal that an agreement at this level of detail between the full observed and constrained simulated lines-of-sight is not utterly expected.\\

\begin{table}
\centering
\begin{tabular}{l@{ }@{ }cccc}
\hline
Cluster &\multicolumn{2}{c}{CLONE/CF2}  &  \multicolumn{2}{c}{CLONE/CF3} \\
Cylinder radius & 10 Mpc & 2.5 Mpc & 10 Mpc & 2.5 Mpc \\
\hline
Virgo     & 0.73 & 0.73 & 0.73 & 0.71 \\  
Centaurus & 0.81 & 0.89 & 0.86 & 0.94 \\
Abell 569 & 0.83  & 0.99 & 0.50  & 0.56  \\
Coma      & 0.84  & 0.96  & 0.84   & 0.96   \\
Abell 85  & 1.0  & 1.0  & 1.0  & 1.0   \\ 
Abell 2256 & 1.0 & 1.0  & 1.0   & 1.0   \\
PGC 765572 & 0.63 & 0.63 & 0.99  &  0.92   \\
PGC 999654 & 1.0  & 1.0 & 1.0   & 1.0  \\ 
PGC 340526 & 0.75  & 0.75 & 1.0   & 1.0   \\ 
PGC 46604 & 1.0   & 1.0 & 1.0  & 1.0  \\
\hline
\end{tabular}
\caption{Kolmogorov-Smirnov statistic or highest distance between the cumulative distribution functions of the observed and simulated lines-of-sight. The latter include the velocity waves.}
\label{tbl:KsStat}
\end{table}

\begin{table}
\centering
\begin{tabular}{l@{ }@{ }ccccc}
\hline
Cluster &\multicolumn{2}{c}{CLONE/CF2}  &  \multicolumn{2}{c}{CLONE/CF3} & Random/CF3 \\
Cylinder & 10 & 2.5 & 10 & 2.5 & 2.5 \\
radius &  Mpc&  Mpc&  Mpc& Mpc & Mpc \\
\hline
Virgo     &  163 &  304  & 157 & 304  & 11200 \\  
Centaurus & 173  & 348 & 171  & 343  & 11221\\
Abell 569 & 294 & 1998 & 155 & 1392 & 7272\\
Coma      & 275   &  722 & 485    & 820 & 6276 \\
Abell 85  & 386 & 727  & 718 & 1082 & 2392\\ 
Abell 2256 & 1453   & 1454 & 1491  & 1491  & 2882\\
PGC 765572 & 1137  &1137 & 1066& 1067 & 2447 \\
PGC 999654 & 747  & 747 & 685  & 685 & 2769 \\ 
PGC 340526 & 875 & 907  & 704 & 736  & 1728\\ 
PGC 46604 & 1341 & 1341 & 1341 & 1341  & 2800 \\
\hline
\end{tabular}
\caption{$\zeta$-metric in km~s$^{-1}$. It measures the difference between the simulated and observed lines-of-sight. The higher $\zeta$ is the more different the lines-of-sight are. The last column gives the average $\zeta$ value when comparing 1,100 random simulated lines-of-sight and the observed ones. The random lines-of-sight were selected to contain at least one halo more massive than 10$^{14}$~M$_\odot$. See the text for a detailed explanation.}
\label{tbl:MetricStat}
\end{table}

To quantify the agreement between simulated and observed lines-of-sight, we use a 2D-Kolmogorov-Smirnov (KS) statistic test applied to the simulated and observed galaxy velocity and position samples. The 2D-KS is an extension of the non-parametric KS test to two-dimensional probability distributions \citep{1983MNRAS.202..615P,1987MNRAS.225..155F}. Like the KS test, it permits comparing samples by quantifying the distance between their distributions. Table \ref{tbl:KsStat} gives the 2D-KS statistic or the highest distance between the cumulative distribution functions of the observed and simulated lines-of-sight that contain the velocity waves. A single 2D-KS statistic value has no particular meaning but several together permit ordering the simulated lines-of-sight from those that match the most their observational counterpart to those that match it the least (smallest to largest values). Virgo line-of-sight happens to be slightly better reproduced than Centaurus' one by the simulation. 2D-KS statistic values are barely different when considering all the subhalos/galaxies within a 10 Mpc radius or solely those within a 2.5 Mpc radius from the line-of-sight. The agreement is comparable when using galaxies from the second catalog (CF2) of the Cosmicflows project or those of the third one. Given that the third catalog has more points and smaller uncertainties, it is encouraging that the values are comparable. Indeed, if the third catalog is closer to the true signal (smaller uncertainties) than the second catalog, the simulation, whose goal is to reproduce the true signal, should tend toward the former. The 2D-KS statistic test cannot, however, take into account observational uncertainties. Finally, 2D-KS statistic values do not differ when using H$_0$~=~67.77 rather than 74~\kms~Mpc$^{-1}$. \\

\begin{figure*}
\center
\vspace{-1.1cm}
\hspace{1cm}\includegraphics[width=0.45  \textwidth, trim={1cm 0 1cm 0},clip]{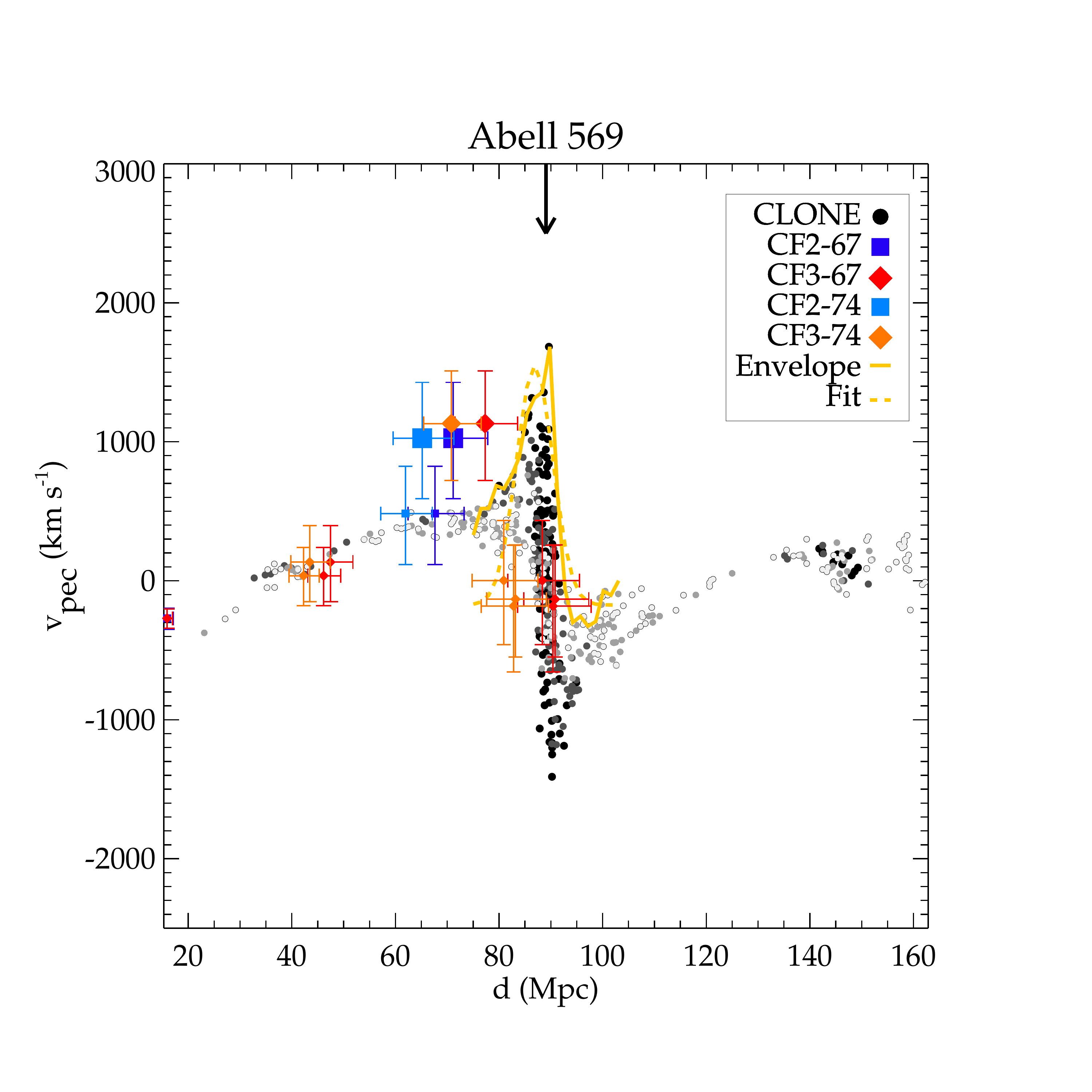}\hspace{-0.5cm}
\includegraphics[width=0.45  \textwidth, trim={1cm 0 1cm 0},clip]{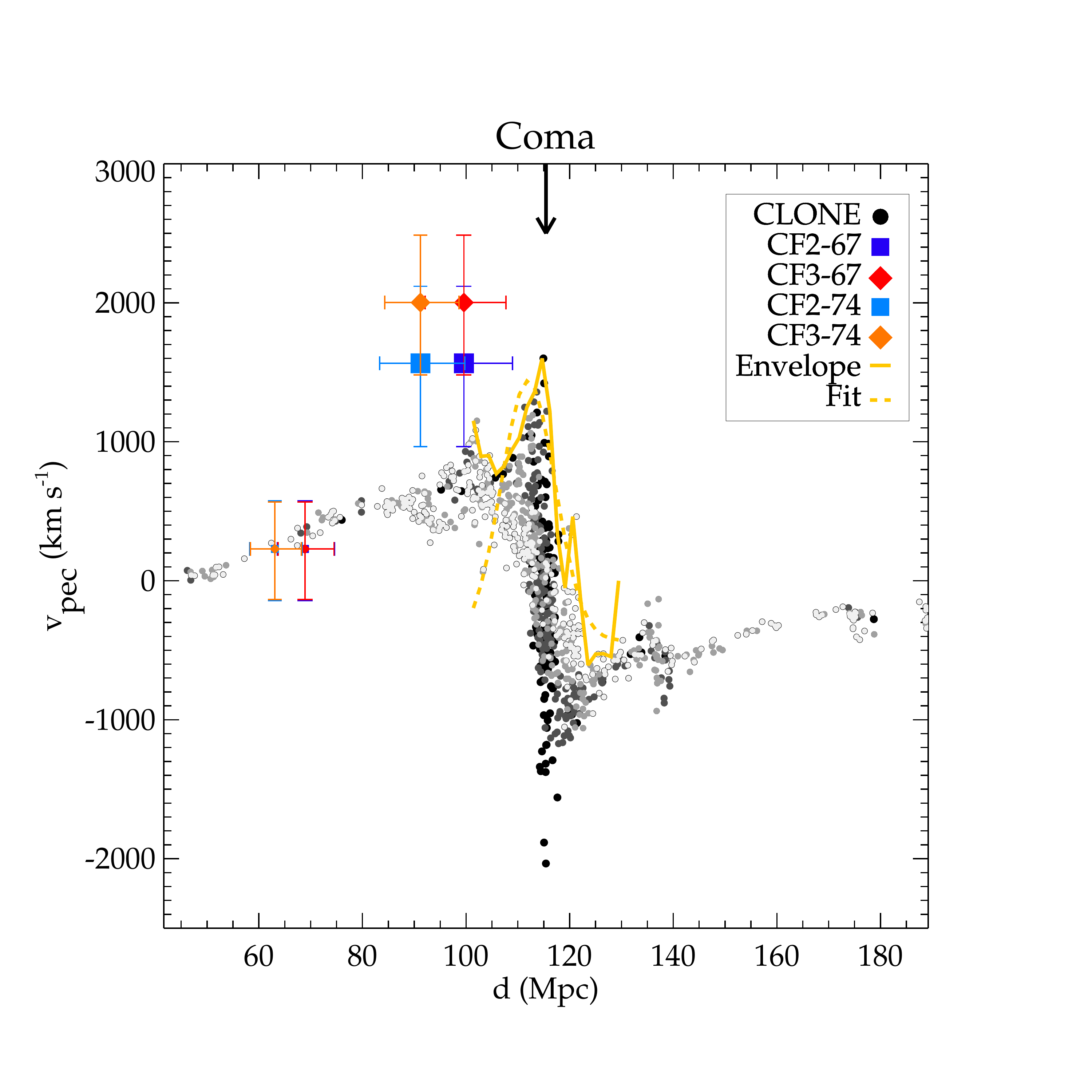}

\vspace{-1.2cm}
\hspace{1cm}\includegraphics[width=0.45 \textwidth, trim={1cm 0 1cm 0},clip]{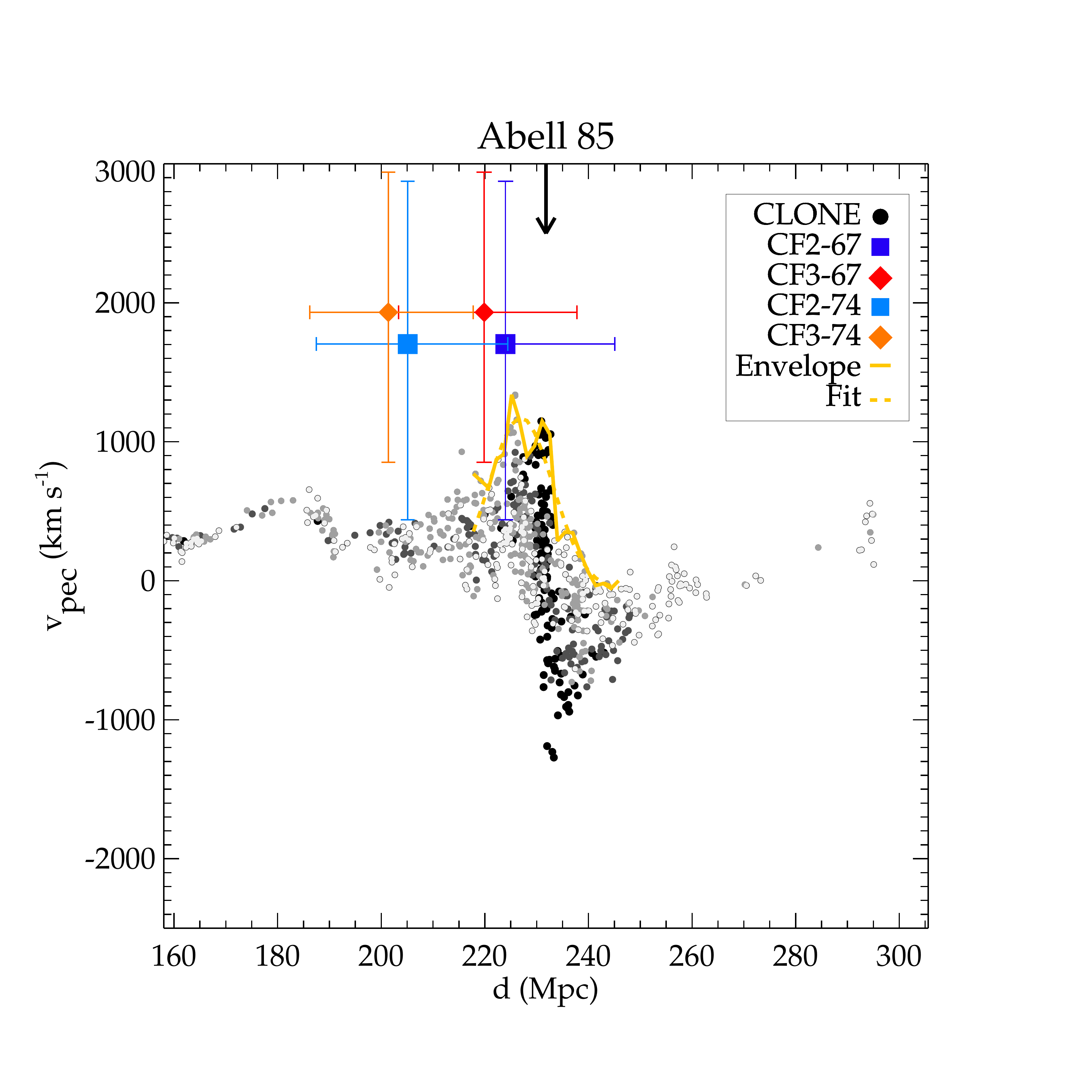}\hspace{-0.5cm}
\includegraphics[width=0.45 \textwidth, trim={1cm 0 1cm 0},clip]{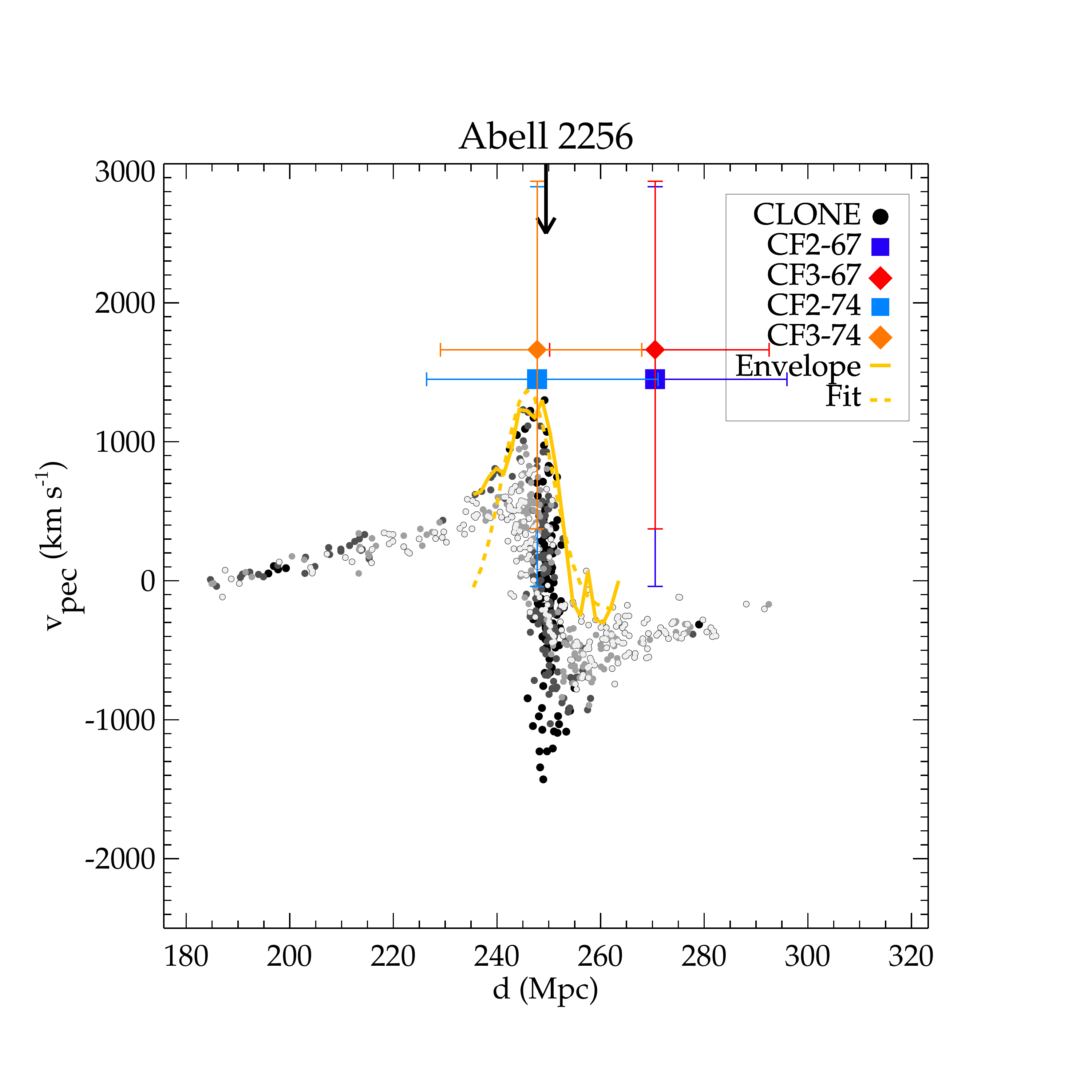}
\vspace{-0.7cm}
\caption{Same as Figure \ref{fig:waves} bottom panels for four clusters at increasing distance from us from left to right, top to bottom. Although these clusters are less constrained, the agreement between observed and simulated waves is still visually good especially for the first two. The dark matter halo masses in the simulation are M=9.0$\times$10$^{14}$M$_\odot$, M=12.6$\times$10$^{14}$M$_\odot$, M=6.6$\times$10$^{14}$M$_\odot$ and M=11.7$\times$10$^{14}$M$_\odot$  for Abell 569, Coma, Abell 85 and Abell 2256 cluster counterparts respectively.}
\label{fig:wavesone}
\end{figure*}

The 2D-KS statistic test cannot take into account the real distance of galaxies. Schematically, it compares only a combination of the four cumulative distributions of galaxies along the lines-of-sight obtained using four directions for counting (smallest to largest distances to the y-axis and vice versa - there is no distance to the observer - , smallest to largest distances to the x-axis - in that case velocities because they are centered on zero - and vice versa). Consequently, we also define our own $\zeta$-metric to compare simulated and observed lines-of-sight as follows:
\begin{equation}
\zeta=\frac{1}{n}\sum^n_{i=1} \rm{min}\{\sqrt{[v_{obs}[i]-v_{sim}]^2+[(d_{obs}[i]-d_{sim})\times H_0]^2}\,\}
\end{equation}
where n is the number of observed galaxies in the line-of-sight. $v_{X}$ are the galaxy/subhalo observed and simulated peculiar velocities and $d_{X}$ are their distances.

Table \ref{tbl:MetricStat} gives the values of $\zeta$ for the different lines-of-sight. $\zeta$-values are only slightly modified by a few percent when changing H$_0$ value. The numbers are slightly in favor of 67.77~\kms~Mpc$^{-1}$ though. Consequently, those for 74~\kms~Mpc$^{-1}$ are reported in the table and should be seen as an upper limit.  Like for the 2D-KS statistic values, $\zeta$-values permit ordering the simulated lines-of-sight (including waves) that are the best reproduction of the observed ones to those that reproduce them the less. {Our $\zeta$-metric seems appropriate as it results in conclusions similar to those obtained with the 2D-KS statistic. However, unlike the 2D-KS statistic, the $\zeta$-metric is sensitive to the real distance of the cluster. It includes differences due to both a difference in height and a shift in position along the entire line-of-sight. A test consists of drawing random lines-of-sight that each include one of the 1,100 most massive halos (minimum mass $\sim$10$^{14}$M$_\odot$) in the simulation. There are then compared to the observed lines-of-sight that include the local clusters given in Table \ref{tbl:MetricStat}. The resulting average $\zeta$-values are given in the last column of Table \ref{tbl:MetricStat}. Values are at least about two to ten times larger than when comparing observed lines-of-sight and their replica in CLONE. The $\zeta$-metric still does not take into account uncertainties. A different metric would be needed to include asymmetric uncertainties on velocities and distances that are lognormal distributed.}\\

\begin{figure*}
\vspace{-0.4cm}
\center
\hspace{-0.2cm}\includegraphics[width=0.29 \textwidth]{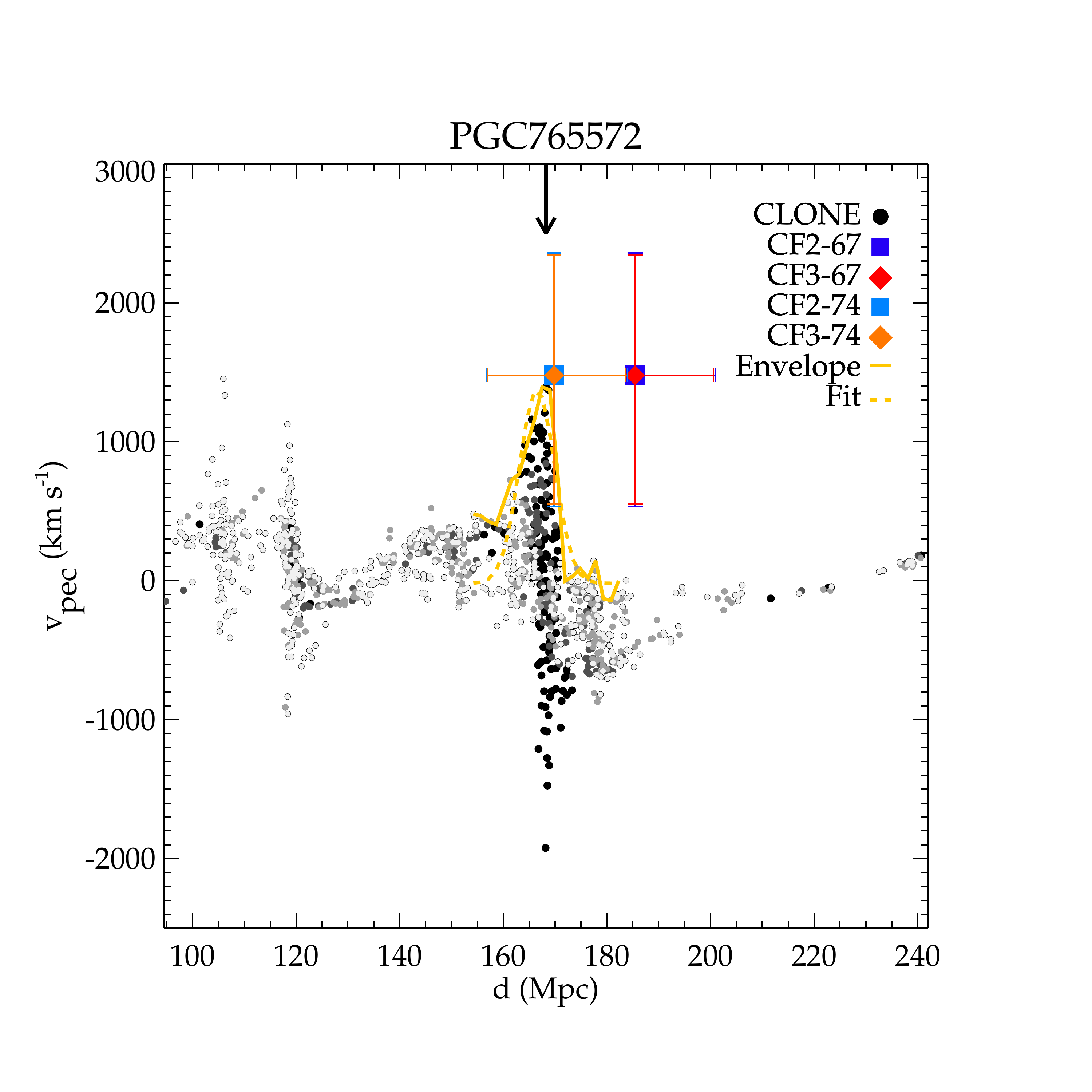}\hspace{-1cm}
\includegraphics[width=0.29 \textwidth]{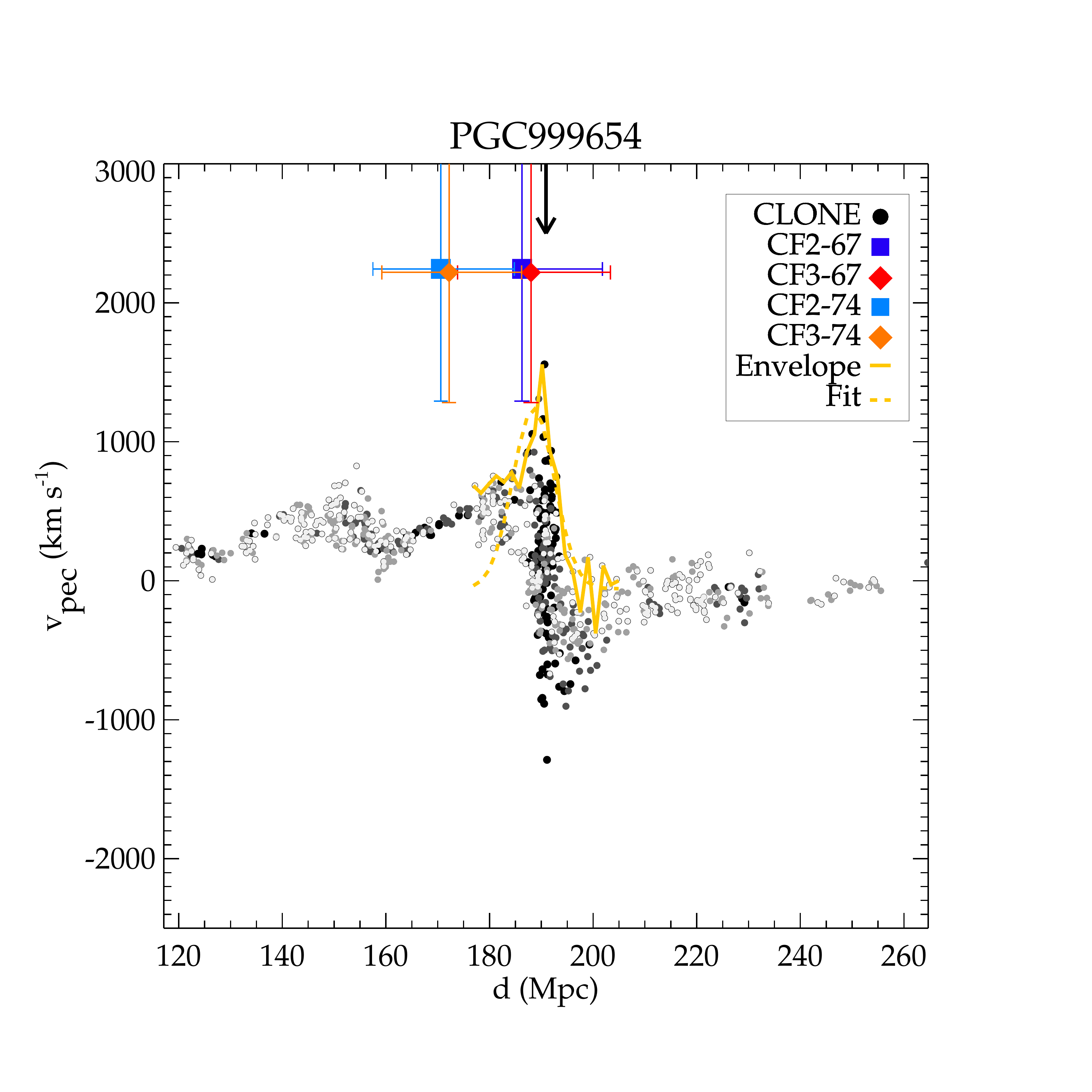}\hspace{-1cm}
\includegraphics[width=0.29 \textwidth]{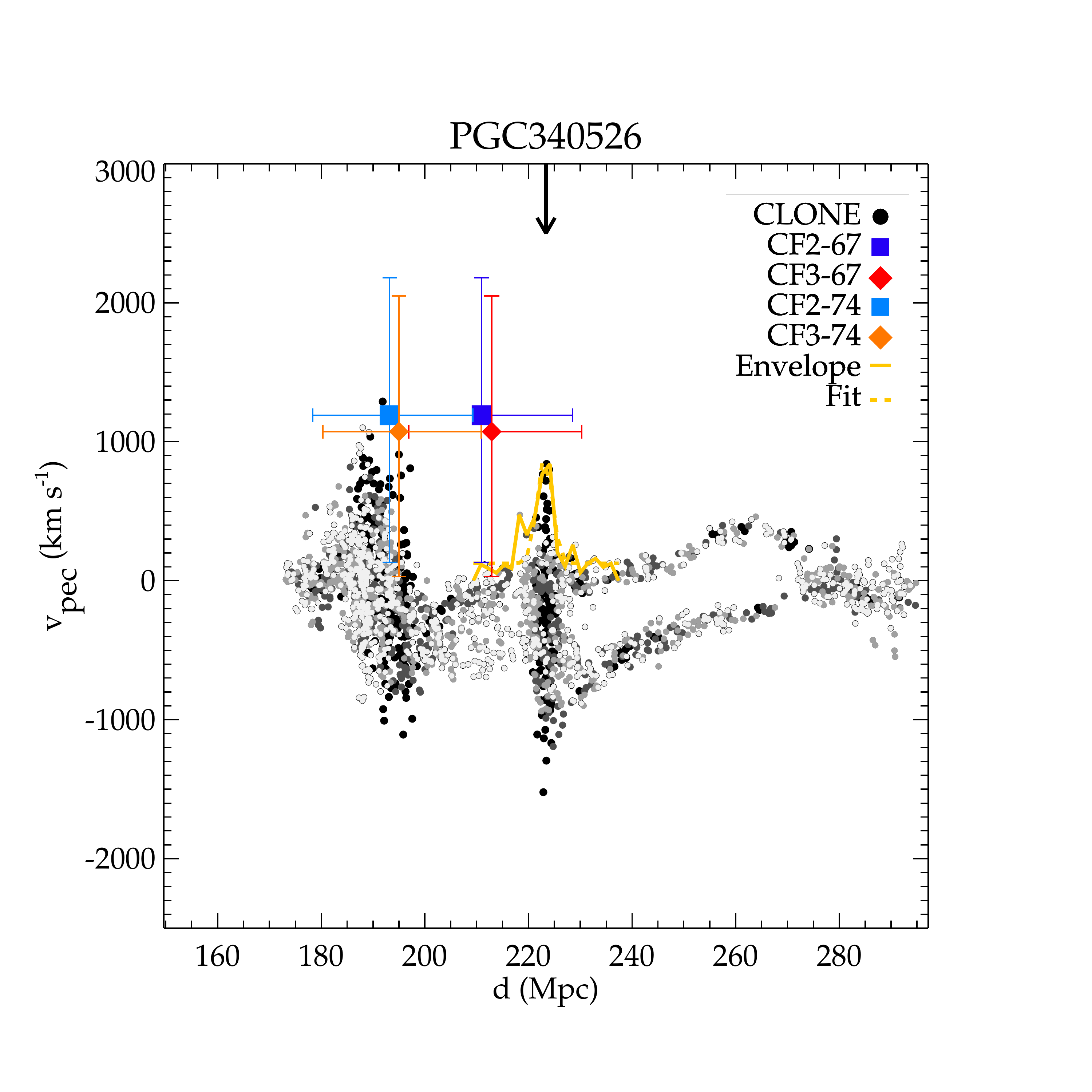}\hspace{-1cm}
\includegraphics[width=0.29 \textwidth]{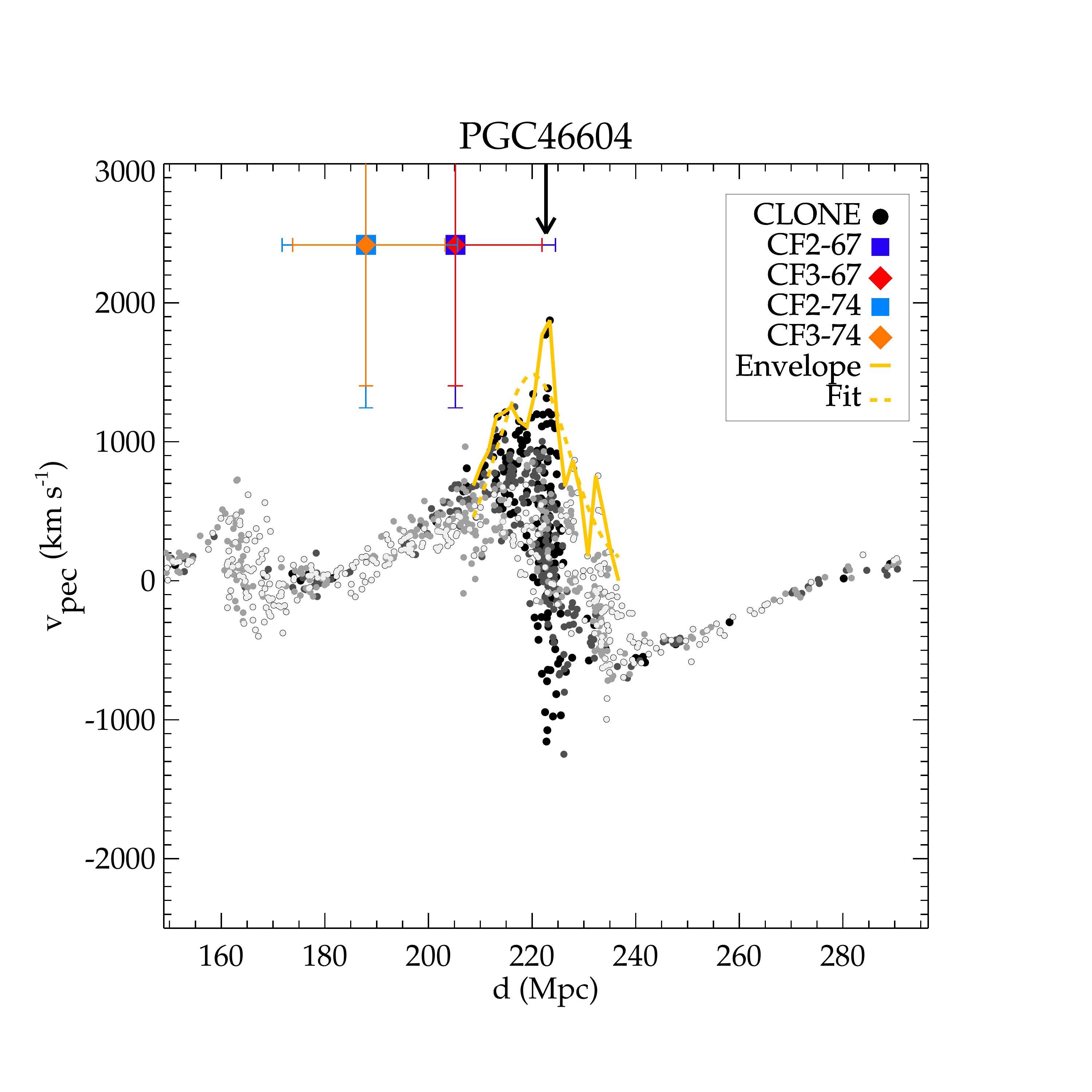}
\vspace{-0.7cm}

\caption{Same as Figure \ref{fig:waves} bottom panels for four additional  clusters. Names at the top of each panel are PGC (Principal Galaxy Catalog) numbers of the galaxies with the highest velocity in the observational catalog at the given locations.}
\label{fig:wavesbis}
\end{figure*}

In the rest of the paper, we work solely with the background expansion subtracted since it does not affect our conclusion and ease the comparisons, studies and analyses. 

Although, by construction, the simulation matches best the local large-scale structure in the inner part, where most of the constraints are, Fig. \ref{fig:wavesone} shows an additional four massive halos that are more distant. These halos are still in agreement with observational clusters that are further away. Tables \ref{tbl:KsStat} and \ref{tbl:MetricStat} confirm the visual impression. The values obtained for both metrics also show the metric limitations and confirm their complementarity. On the one hand, the $\zeta$-metric is more robust to small samples than the 2D-KS statistic. {For example, Abell 569 has a smaller observational sample in the second catalog of the Cosmicflows project than in the third one. The $\zeta$-values when comparing both observational samples to the simulated one differ by only a few percent. The 2D-KS statistic values grandly differ. One the other hand, the 2D-KS statistic is more robust to observational uncertainties. Given their uncertainties, the peculiar velocity values of galaxies in Coma and Abell 85's surroundings are compatible between the second and third catalogs of the Cosmicflows project. They are higher though in the third catalog. This results in higher $\zeta$-metric values when comparing lines-of-sight from this third catalog to the simulated one rather than those from the second catalog to the simulated one. It is not completely unexpected that the simulated lines-of-sight match better those from the second catalog than the third one at a fixed number of datapoints and similar uncertainties.} Indeed, the second catalog is the starting point to build the constrained initial conditions. More precisely, peculiar velocity values of groups and field galaxies in this area at z=0 are used to estimate their progenitors' position and velocity through the reconstruction of the displacement/velocity fields. These progenitors' position and velocity are then used to constrain the initial conditions. Initial conditions are then evolved. There is finally a (non-linear) relation between the final wave and the initial constraints. There is, however, no convolution as non-field galaxy peculiar velocities are not used to directly and individually constrain the initial conditions that are, moreover, evolved down to z=0.\\
 
Since observed galaxies with low distance uncertainties are usually not exactly along the line-of-sight of the massive clusters, their velocity constitutes a lower limit for the mass estimate of the observed clusters. Indeed, galaxies perfectly aligned with the observer and the cluster would have the highest possible velocity but such galaxies are difficult to distinguish from those belonging to the cluster. Consequently, for Virgo, Centaurus and Abell 569, the maximum peculiar velocity in the simulation is slightly higher than that in the observations. It confirms that the simulated clusters have reached the low mass limit set by the observations. Moreover, the difference between the observed and simulated wave maxima is small enough that masses are within the same mass range according to the Least Action modeling \citep[see for instance][]{2005ApJ...635L.113M,2004ogci.conf..205T}. This agreement is confirmed by observational data that follow the wave shape so as to reproduce its width. The next section expands on the link between wave properties and cluster masses. Note the adequacy between simulated and observed velocity wave shapes for Abell 569: small uncertainty peculiar velocities, not used to constrain this wave progenitor in the \emph{initial conditions' linear} regime, follow the simulated wave contour. There are indeed two orange/red datapoints from the third catalog that have no blue counterpart in the second catalog. The 2D-KS statistic small value confirms the adequacy.

Given their hosted galaxy peculiar velocity uncertainties, the lower mass limits of Coma, Abell~85 and Abell 2256 are reached. This is not fully expected given that these clusters are at the edge of the constrained region (50\%, 90\% and 99\% of the constraints are in $\sim$75-80, 150-160 and 275-290~Mpc). Additional precise observational data are, however, required to probe the wave slopes and check their width to eventually tighten the constraint on the masses.\\

Fig. \ref{fig:wavesbis} shows four additional velocity waves born from massive dark matter halos to which we can associate observed galaxies. The galaxies with the largest peculiar velocities are identified by their PGC (Principal Galaxy Catalog) number at the top of each panel. Given the distance of these clusters and the sparsity and limit of our constraint-catalog, there is an agreement. Tables \ref{tbl:KsStat} and \ref{tbl:MetricStat} confirm again the visual impression. They also highlight again the limitations of both metrics. Both values must be given together to conclude on how much the observed and simulated lines-of-sight match. We identify other simulated velocity waves corresponding to local clusters (e.g. in the Perseus-Pisces region) but observational data is not of sufficient quality or absent in the infall region for comparisons. Nonetheless, all the halos and associated waves are used for the studies in the next section. The mass range is actually extended down to 2~10$^{14}$M$_\odot$.

\section{Wave properties versus cluster masses}

In the following, cluster and galaxy terminologies are used in place of the (massive) dark matter halo one.
\subsection{Amplitude}

\begin{figure}
\vspace{-0.9cm}
\flushleft
\hspace{-0.7cm}\includegraphics[width=0.58 \textwidth,trim={1cm 0 0 0},clip]{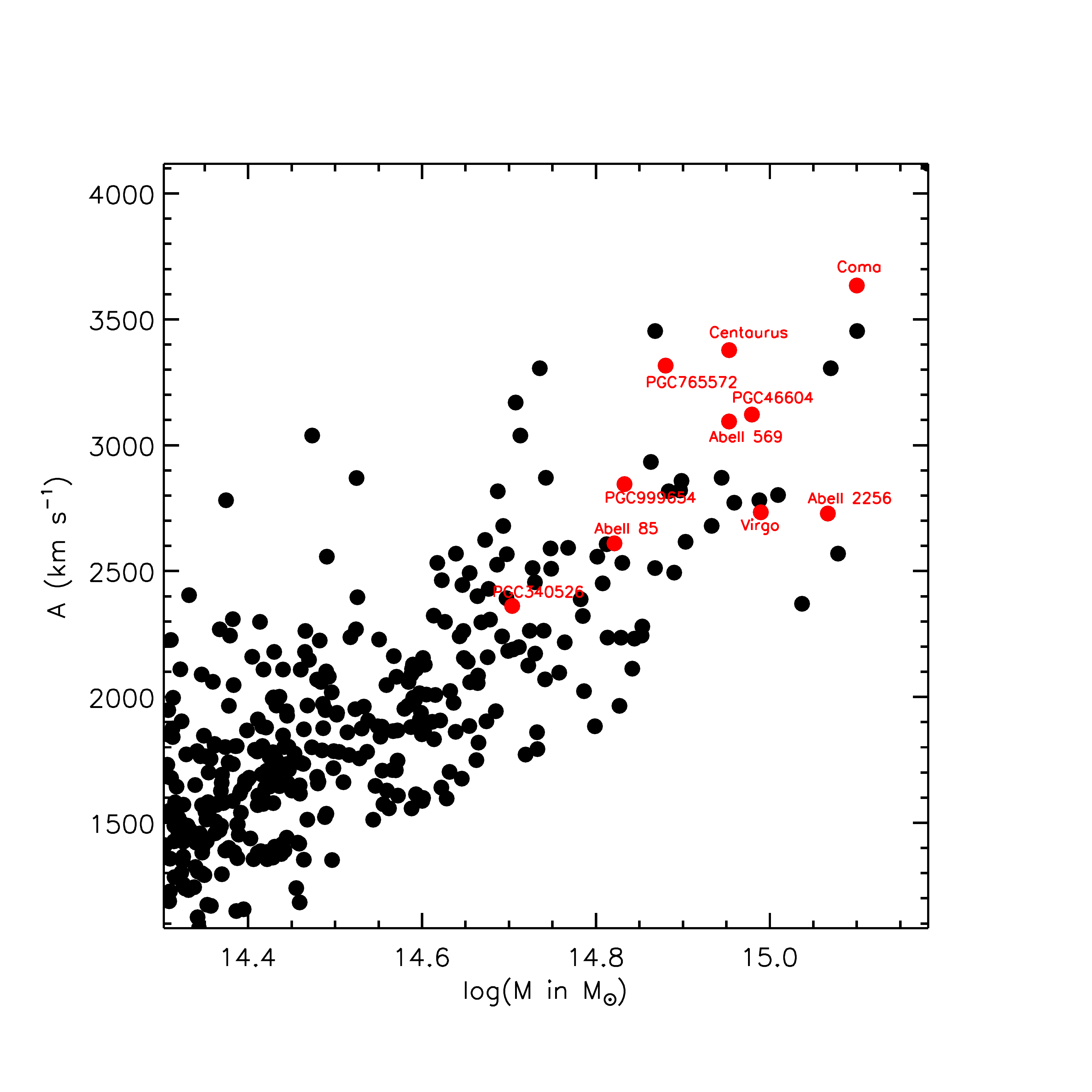}
\vspace{-1cm}

\caption{Amplitude of the simulated velocity waves as a function of the dark matter halo masses. Amplitudes are obtained with dark matter halo minimum and maximum velocities. Halos shown in Fig. \ref{fig:waves} to \ref{fig:wavesbis} are identified in red.}
\label{fig:amplitude}
\end{figure}

The wave amplitude is the first obvious property to check against halo mass. Indeed, the deeper the gravitational potential well is, the faster galaxies should fall onto it. The amplitude is defined as the difference between the maximum and minimum peculiar velocities of galaxies falling onto the cluster either from the front or from behind with respect to the synthetic observer.  Fig. \ref{fig:amplitude} shows the amplitude of the simulated velocity waves as a function of the dark matter halo masses. Each filled black and red circle corresponds to a halo. Red ones stand for clusters identified in Fig. \ref{fig:waves} to \ref{fig:wavesbis}. While it is immediate to notice that there is a correlation between the wave amplitude and the halo mass, one can also point out that the amplitude is extremely difficult to measure in observational data and that there is a residual scatter. Indeed, measuring the amplitude in observational data implies getting accurate distance (peculiar velocity) estimates of galaxies exactly in the line-of-sight of the cluster with respect to us. It supposes first that there are actually galaxies exactly aligned. Then, identifying these galaxies and measuring their distances with high accuracy, while they fall onto the cluster from the front is already quite a challenge, let alone when they fall from behind. 

In any case, the residual scatter suggests that the amplitude, be it measurable, alone cannot be used as a precise proxy for cluster mass estimates. Part of this scatter is probably due to the fact the galaxies are not perfectly aligned with us and the cluster. The gravitational potential well shape might also be responsible for another part of this scatter. To a lesser extent, the large-scale structure environment might also play a role. 

\subsection{Height}

\begin{figure}
\vspace{-0.9cm}
\flushleft
\includegraphics[width=0.58 \textwidth, trim={1.2cm 0 0 0},clip]{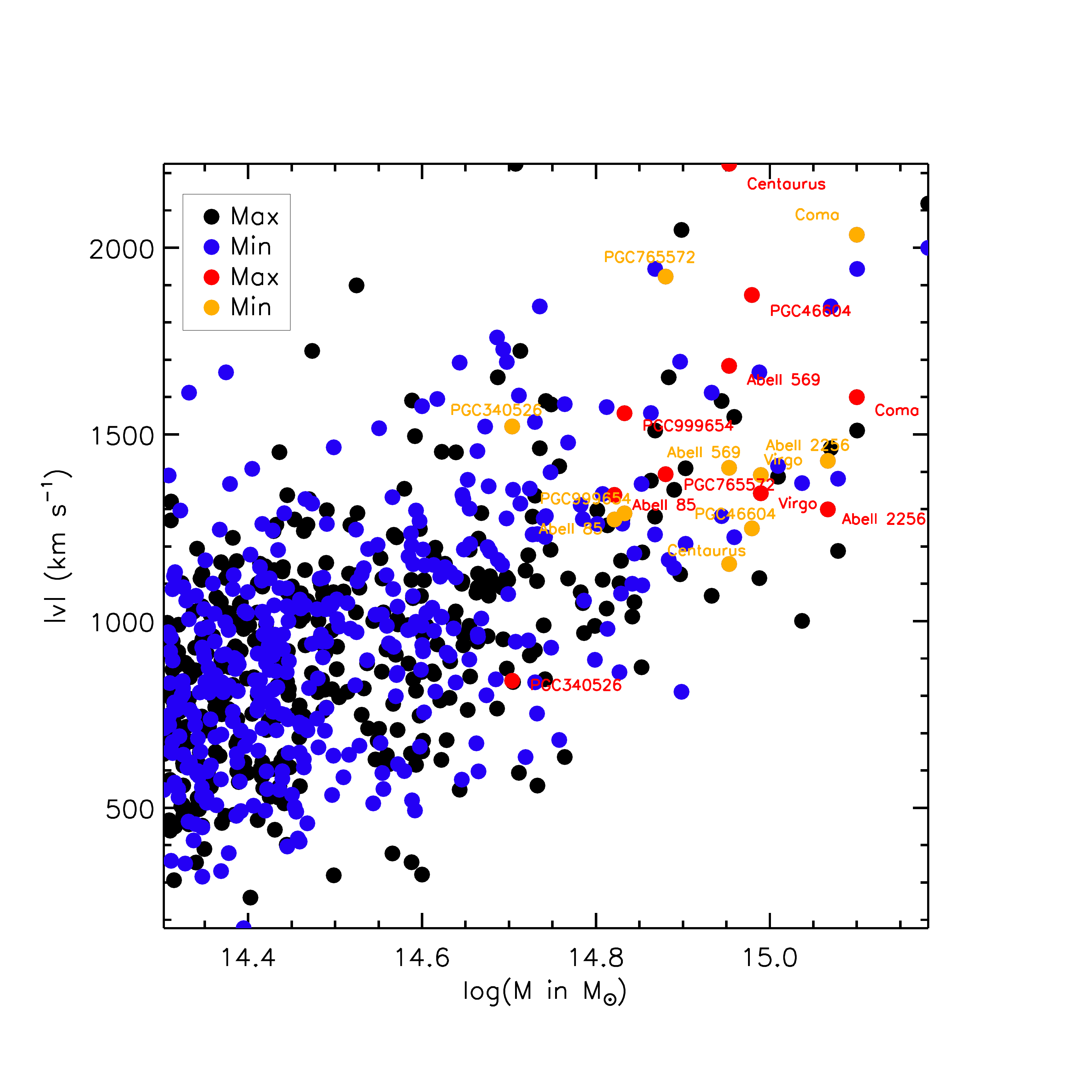}
\vspace{-1cm}

\caption{Dark (blue) filled circles are heights of the simulated positive(negative)-half velocity waves as a function of the dark matter halo masses. Heights corresponds to maximum (minimum) dark matter halo velocities. Halos shown in Fig. \ref{fig:waves} to \ref{fig:wavesbis} are identified in red (orange).}
\label{fig:height}
\end{figure}

While the wave height is not expected to be a better proxy than the wave amplitude, it is interesting to check whether there still is a tight enough correlation. Indeed, while it is challenging to have precise distance measurements for both galaxies falling from the front and from behind a cluster in the line-of-sight with respect to us, it might be feasible especially for galaxies falling from the front. The height is thus defined as the maximum (minimum) peculiar velocities of galaxies falling onto the cluster from the front (behind) with respect to the synthetic observer. In Fig. \ref{fig:height}, each black and red (blue and orange) filled circle stands for the height of a dark matter halo positive-(negative-)half wave as a function of its mass. Heights are measured with velocities of galaxies falling from the front (behind) into the clusters. Red and orange are used for dark matter halos from Fig. \ref{fig:waves} to \ref{fig:wavesbis}.

A similar correlation as with the amplitude is found although with a somewhat larger scatter. Interestingly it also shows that velocity waves are not symmetric: their maximum differs from their minimum. Both the potential well shape and the non-perfect alignment observer-galaxy-cluster or observer-cluster-galaxy might be the reason for this asymmetry. Nonetheless because there still is a correlation and because in observational data it is easier to get accurate datapoints at the wave front than in its wake, it is legitimate to focus on the positive-half velocity wave shape to study more thoroughly the relation with the halo mass.
 
\subsection{Height, width and continuum}

After deriving the positive-half wave envelope of every dark matter halo, as a proof-of-concept we choose to fit the simplest model possible, a Gaussian-plus-continuum model, to each one of them as follows:
\begin{equation}
v_{pec} = A_{fit} ~\times~ \mathrm{e}^{\frac{-(d-d_0)^2}{2 \sigma_{fit}^2}}+ C_{fit}
\end{equation}
where $A_{fit}$, $\sigma_{fit}$ and $C_{fit}$ are respectively the Gaussian amplitude, its standard deviation and a continuum. $d_0$ depends on the halo distance and has no other purpose than centering the Gaussian on zero. Its sole physical meaning is to be the actual distance of the halo. The amplitude is related to the positive-half wave envelope height while the standard deviation is linked to its width. Finally, the continuum gives the positive-half wave offset from a zero average velocity. For visualization, envelopes and their fits for halos presented in Fig. \ref{fig:waves} to \ref{fig:wavesbis} are shown as solid and dashed lines in these same figures. In observational catalogs, velocity waves are more likely to be highlighted with only a few galaxies. Degrading the information in the simulation by using only velocity-wave envelopes and their fits is, thus, coherent with the information available in observational catalogs for further applications to observations. \\

\begin{figure*}
\vspace{-0.4cm}
\centering
\hspace{-0.5cm}\includegraphics[width=0.8 \textwidth]{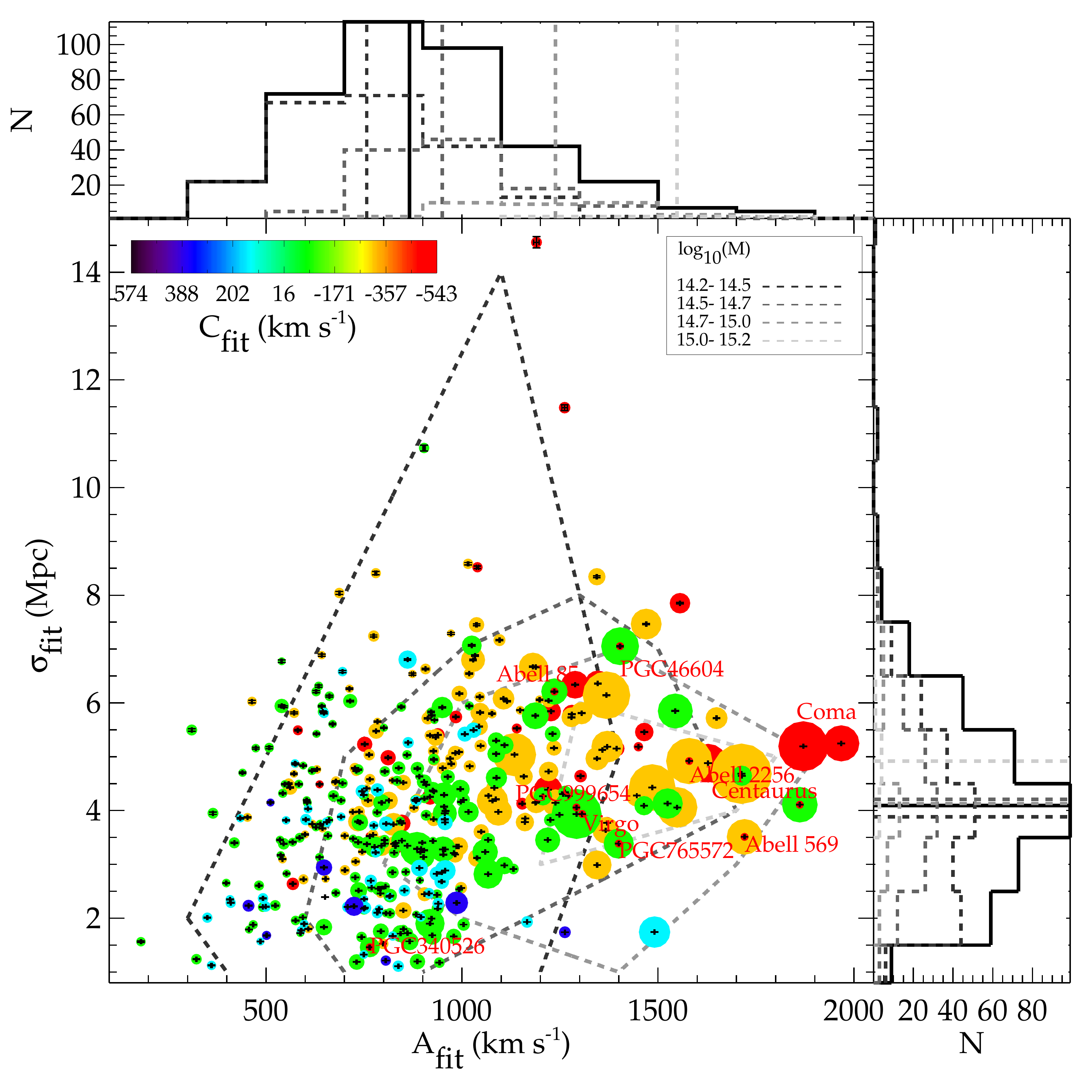}
\vspace{-0.3cm}

\caption{Parameters of the Gaussian-plus-continuum fit to the simulated positive-half velocity waves. $\sigma_{fit}$ stands for the Gaussian standard deviation, $A_{fit}$ for its amplitude and $C_{fit}$ for the continuum. The filled circle sizes are proportional to dark matter halo masses. Tiny error bars on the standard deviation and amplitude resulting from fitting the envelopes highlight the adequacy of the model choice. Halos shown in Fig. \ref{fig:waves} to \ref{fig:wavesbis} are identified with red nametags and additional small filled red circles. {The dashed grey lines stand for the surfaces that encompass 95\% of the halos of a given mass bin. The mass bin increases from dark to light grey. Histograms of the amplitude and standard deviation values for the full sample of halos (solid black line) and per bin of mass (dashed grey lines) are given in the top and right panels respectively together with medians. The more massive the halos are the larger the amplitude is, and to a lesser extent the standard deviation is, on average. The scatter in the standard deviation values increases with the decrease in mass probably because the environment plays a more important role for the halos of the smallest mass bin, while massive halos are more active in shaping their environment.}}
\label{fig:fits}
\end{figure*}

Fig. \ref{fig:fits} gathers the three parameters of the fits and halo masses for a concomitant study to highlight an eventual multi-parameter correlation. The Gaussian amplitude is represented as a function of the Gaussian standard deviation while the color scale stands for the continuum. From black, violet to red, the continuum decreases from positive values to negative ones. The model uncertainty is shown as error bars for the amplitude and standard deviation. The color scale smoothness includes the continuum uncertainty. The Gaussian-plus-continuum model choice proves to be robust given the tiny error bars that it results in. The filled circle sizes are proportional to the dark matter halo masses. Finally, an additional small filled red circle is used to identify each halo analyzed in Fig. \ref{fig:waves} to \ref{fig:wavesbis}.\\

The previous subsection (4.2) showed that there is a correlation between the wave height and the halo mass. It is thus not surprising to find back that the more massive the halo is (larger circle), the larger the Gaussian amplitude is (larger value). {The histograms in the top panel confirm that the median amplitude increases with the bin of mass (dashed lines from dark grey to light grey)}. As stated above, the Gaussian amplitude is the counterpart of the positive-half wave height. 

\begin{figure*}
\center
\vspace{-.4cm}
\includegraphics[width=0.88 \textwidth]{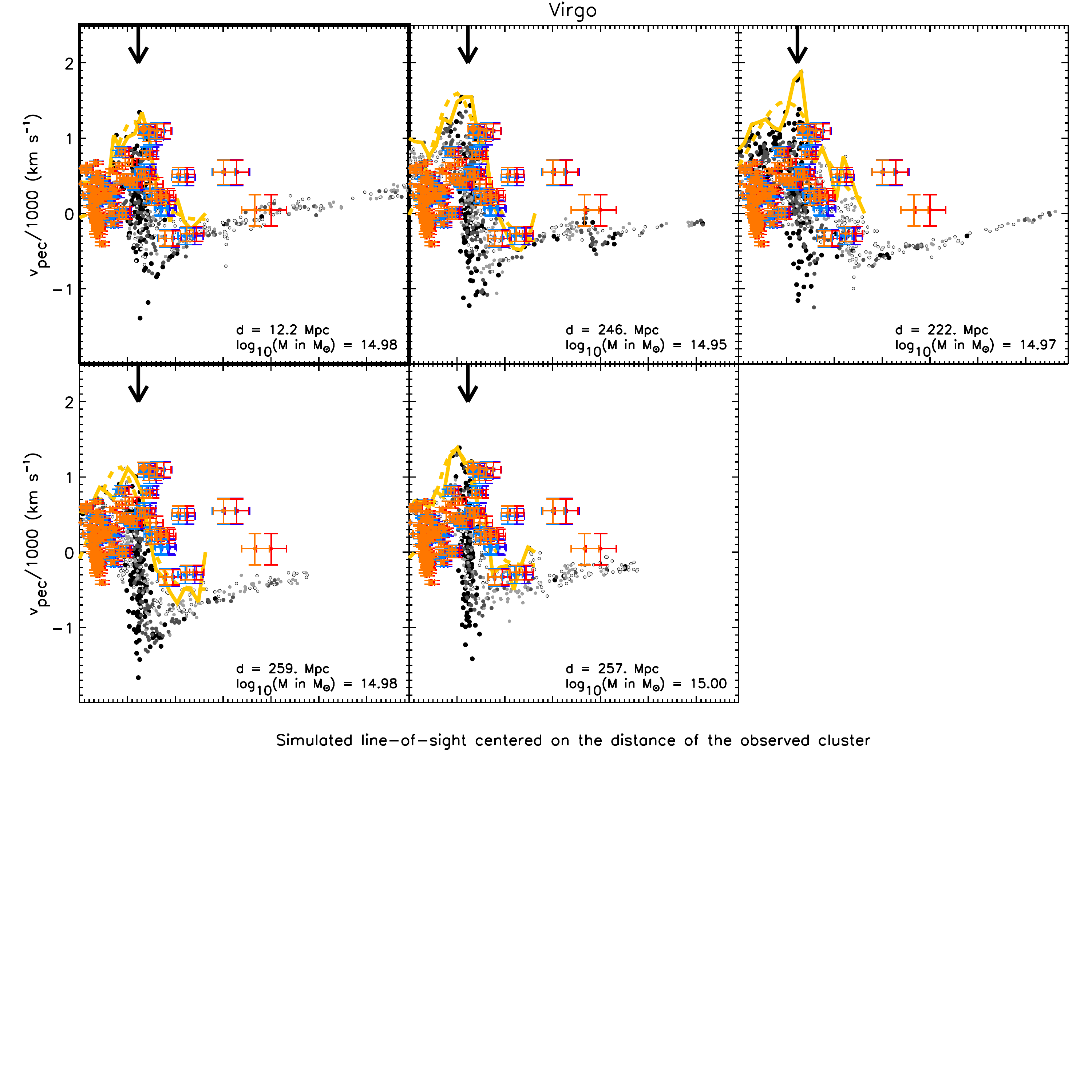}
\vspace{-5cm}

\includegraphics[width=0.88 \textwidth]{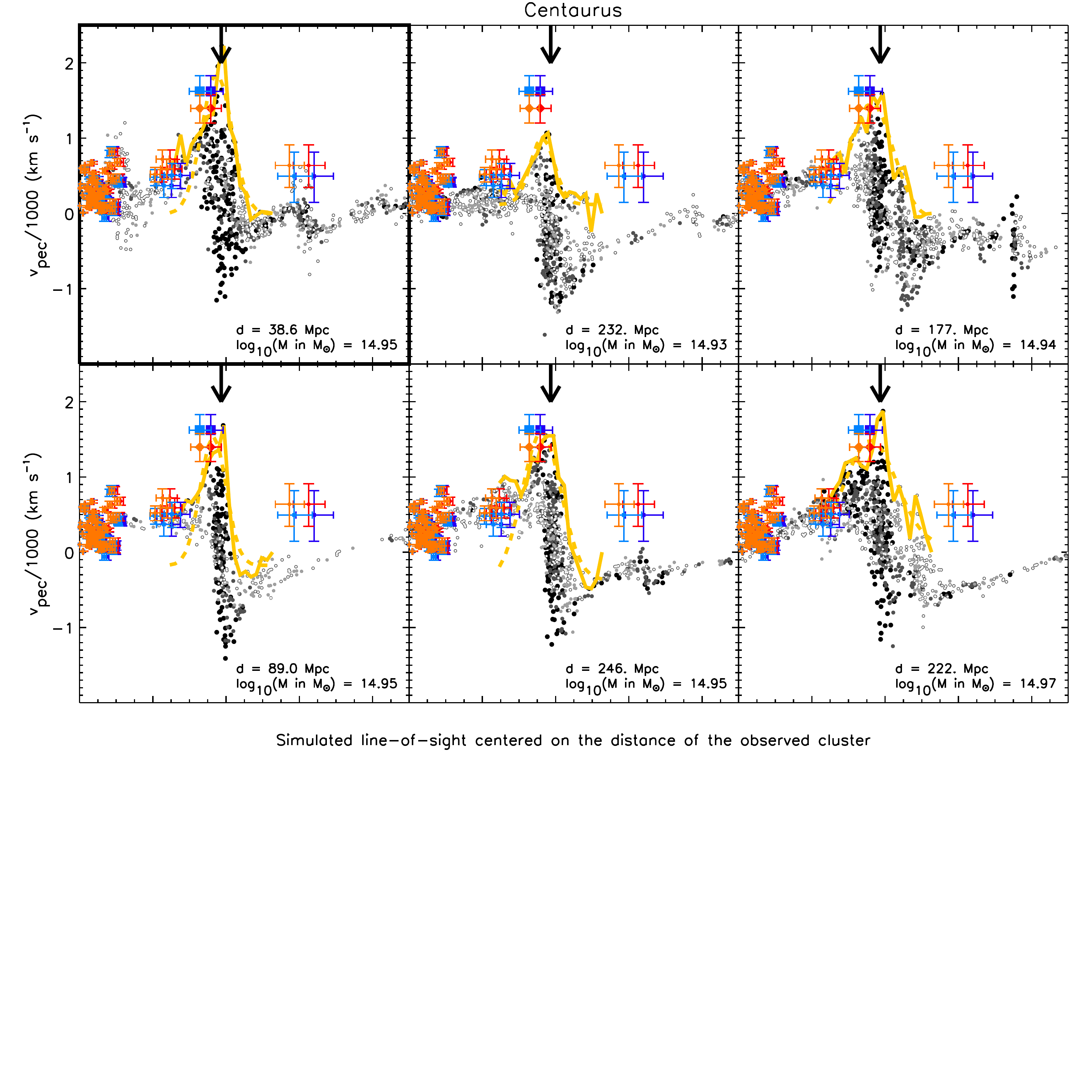}
\vspace{-4.9cm}

\caption{Same as Figure \ref{fig:waves} bottom panels. The first panel of each subfigure stands for the line-of-sight that includes the replica of the observed cluster whose name is given at the top of the panels. The other panels show the lines-of-sight that include halos with very similar masses to the replicas. They have though been shifted so as to be located as the same distance as the simulated replicas in order to be able to overplot observational datapoints. Their original distance to the observer in the simulation is given together with their mass in each panel.}
\label{fig:multiwaves1}
\end{figure*}

In addition, there is a small correlation between the amplitude and standard deviation thus halo mass. More massive halos seem to give birth to wider waves on average. {The histograms in the right panel confirm that the median standard deviation increases with the bin of mass (dashed lines from dark grey to light grey). The scatter however increases with the decrease in mass as shown by the dashed contours in the middle panel. These contours delimit the amplitude-standard deviation region containing 95\% of the clusters of a given mass bin. It highlights the standard deviation dependency on the halo triaxiality and thus on its orientation with respect to us, and to the environment especially for the halos of the lowest mass bin}. 
A similar conclusion is valid for the continuum, the smaller the continuum but for extreme values is, the more massive the halo is on average. The scatter is also quite large in that case. {A correlation between the standard deviation value and that of the continuum is noticeable especially for the low mass bin.} A strong dependence on the global environment of the dark matter halo, in addition to the halo mass, might thus also be a cause here. 

Interestingly a general pattern emerges: \\ 
$\bullet$ {the most massive halos ($\gtrsim$ 5~10$^{14}$~M$_\odot$) tend to give birth to positive-half waves that have a positive or slightly negative continuum in addition to high amplitude and standard deviation values (two lighter grey colors)}.\\
$\bullet$ {the less massive halos  (2~10$^{14}$~M$_\odot$$\lesssim$M$\lesssim$ 3~10$^{14}$~M$_\odot$) tend to give birth to positive-half waves that have low amplitude, and highly scattered but correlated continuum and standard deviation values (darker grey color)}.\\
$\bullet$ {intermediate mass halos (3~10$^{14}$~M$_\odot$$\lesssim$M$\lesssim$ 5~10$^{14}$~M$_\odot$) give rise to positive-half waves that have intermediate amplitude values and on average higher continuum values. Such values permit distinguishing them from the most massive halos with which the can share high amplitude and possibly standard deviation values (intermediate grey color)}.\\
It is probable that the global environment or cosmic web is responsible for such a finding. We will investigate this link in more detail in future studies. \\

Halos are also segregated in different continuum value classes. There seems to be a different correlation for each continuum value class: \\
$\bullet$ Halos with fits resulting in a high (close to zero) continuum value seem to have masses correlated with the Gaussian amplitudes but not so much with the Gaussian standard deviations that appear to have low values (present a large scatter). \\
$\bullet$ Halos with fits resulting in a very low continuum value have both amplitudes and standard deviations correlated together as well as with the masses.\\
$\bullet$ Halos with fits resulting in either positive or negative intermediate continuum values present masses correlated with amplitudes and to a lesser extent with standard deviations.

To summarize, since the fit parameters are interdependent, a global fit to the velocity wave seems the best approach to obtain cluster rough mass estimates rather than single and independent measurements of amplitude, height and width. Because different categories appear among halos, in future studies, a machine learning approach might become handy to actually get accurate enough mass estimates from sparse observations. In a first approach, the simple Gaussian-plus-continuum fit presented here could be used as summary statistics. Meanwhile, in the next subsection, we propose another single value summary statistics: the centered $\zeta$-metric.

\begin{figure*}[h!]
\center
\vspace{-.4cm}

\includegraphics[width=0.88\textwidth]{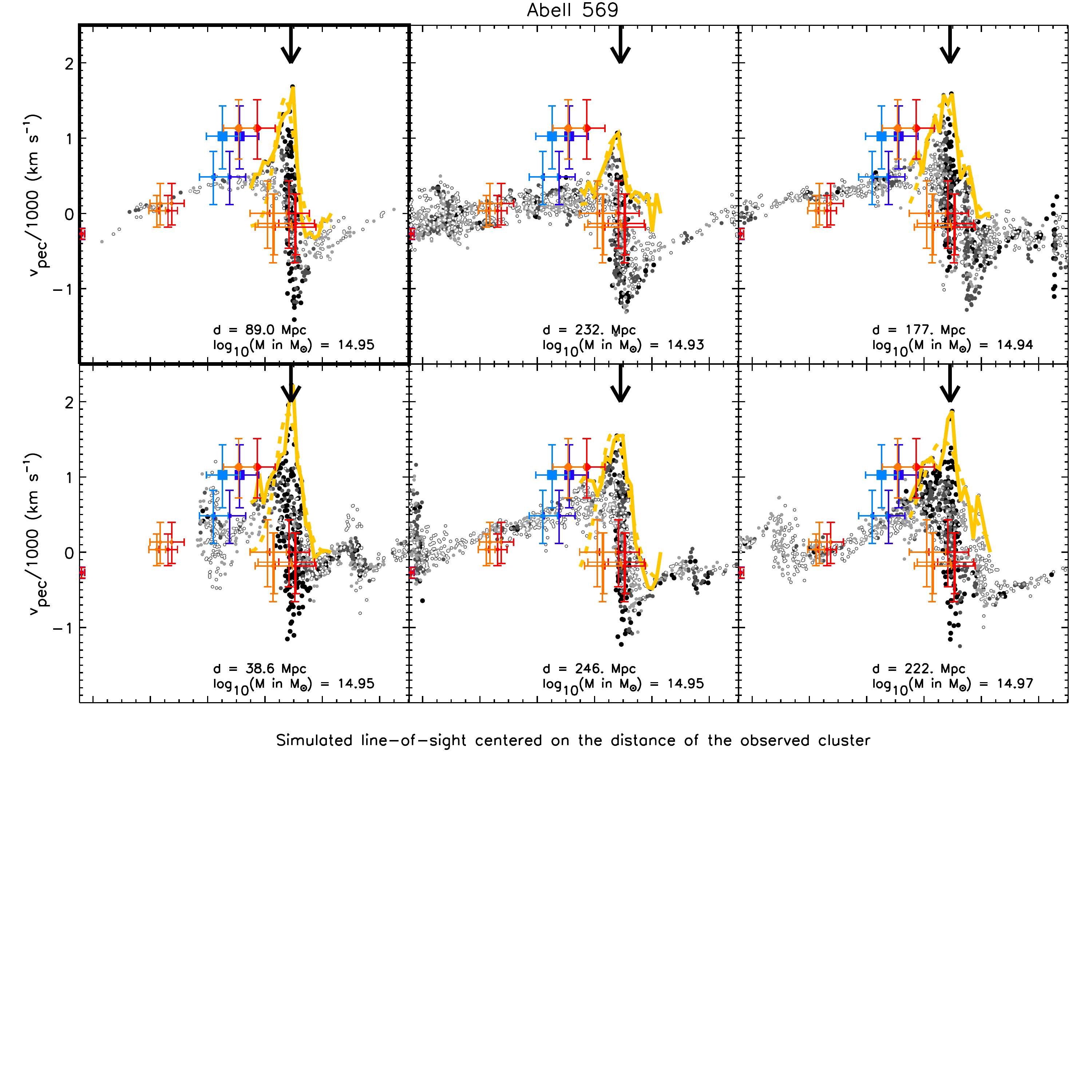}
\vspace{-4.9cm}

\includegraphics[width=0.88\textwidth]{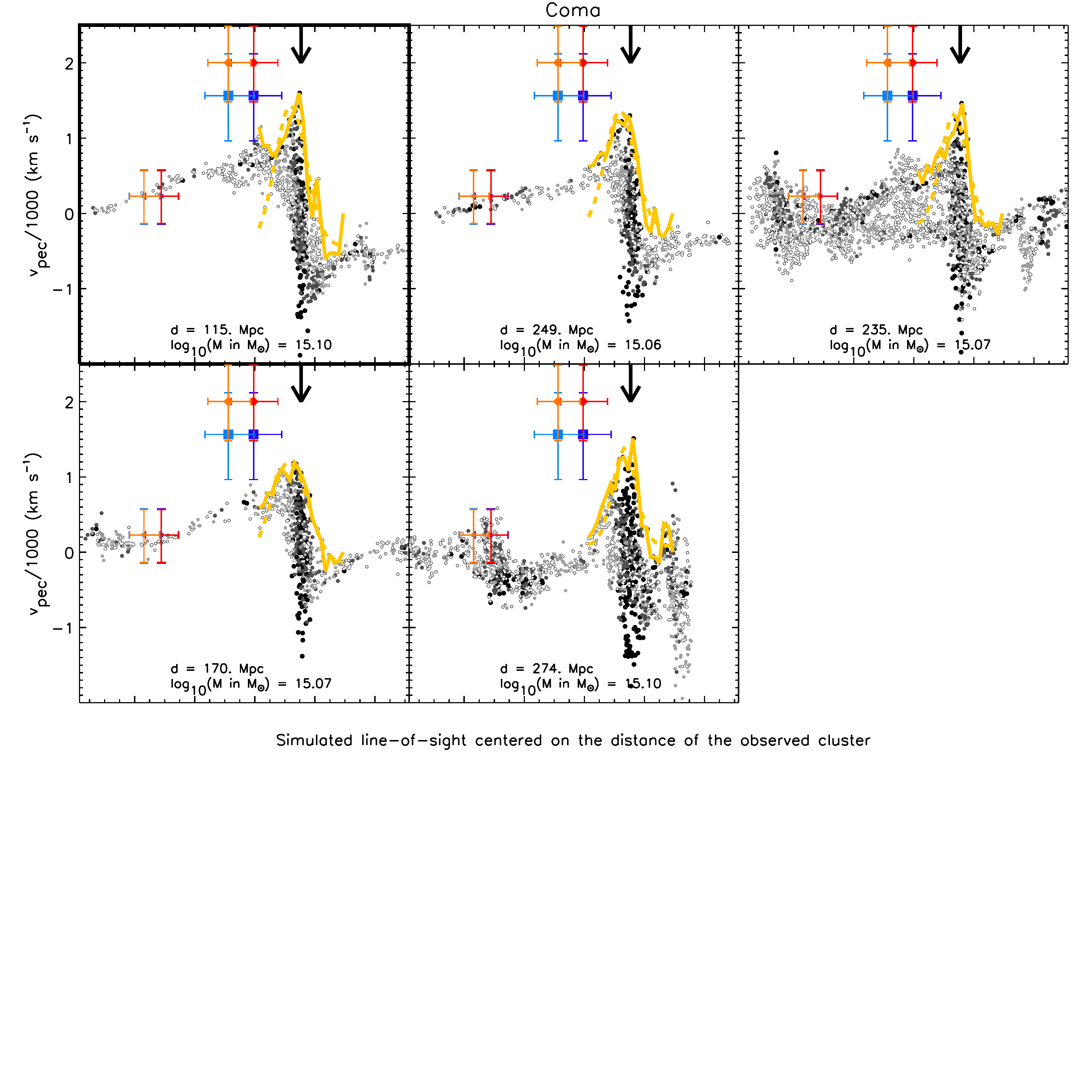}
\vspace{-5cm}

\caption{Same as Figure \ref{fig:multiwaves1} but for the Abell569 and Coma clusters.}
\label{fig:multiwaves2}
\end{figure*}

\subsection{Centered $\zeta$-metric}

{The $\zeta$-metric might be a good proxy for estimating cluster mass, as the wave shape is related to the cluster mass. Since it measures the adequacy between observational datapoints and simulated waves, it could assign the best-fitted wave type, thus mass, to an observational datapoint set. However, the $\zeta$-metric depends on the distance to the cluster. It thus can only be applied to observed clusters and their simulated replicas. To remove this dependence, the centered $\zeta$-metric, or $\zeta_c$-metric is introduced:}

\begin{equation}
\zeta_c=\frac{1}{n}\sum^n_{i=1} \rm{min}\{\sqrt{[v_{obs}[i]-v_{sim}]^2+[(d_{obs}[i]-\tilde d_{sim})\times H_0]^2}\,\}
\end{equation}
where n is still the number of observed galaxies in the line-of-sight. $v_{X}$ are again the galaxy/subhalo observed and simulated peculiar velocities and $d_{X}$ are their distances. However, the distances of the simulated subhalos, $d_{sim}$, are shifted so that simulated and observed clusters' centers match, $\tilde d_{sim}$. \\

{In the previous sections, four simulated and observed clusters are matched using more than one observational datapoint that passed the small uncertainty threshold. These clusters are Virgo, Centaurus, Abell 569 and Coma. The position of other halos are re-centered to appear at the same distance as the replicas. Figures \ref{fig:multiwaves1} and \ref{fig:multiwaves2} show the lines-of-sight, waves included, for the re-centered halos the closest in mass to the different replicas. Observational datapoints corresponding to the replicas' counterparts are overplotted. The top left panels show the line-of-sight of the replicas. The other panels highlight those of the other halos, i.e. sharing similar masses but different environment and history. Overall, the lines-of-sight hosting halos of similar masses appear in agreement with the observational datapoints they have been assigned to. This adequacy reinforces the existence of a strong link between velocity waves and cluster masses. }

\begin{figure*}
\center
\vspace{-1.0cm}
\hspace{-0.5cm}\includegraphics[width=0.5\textwidth]{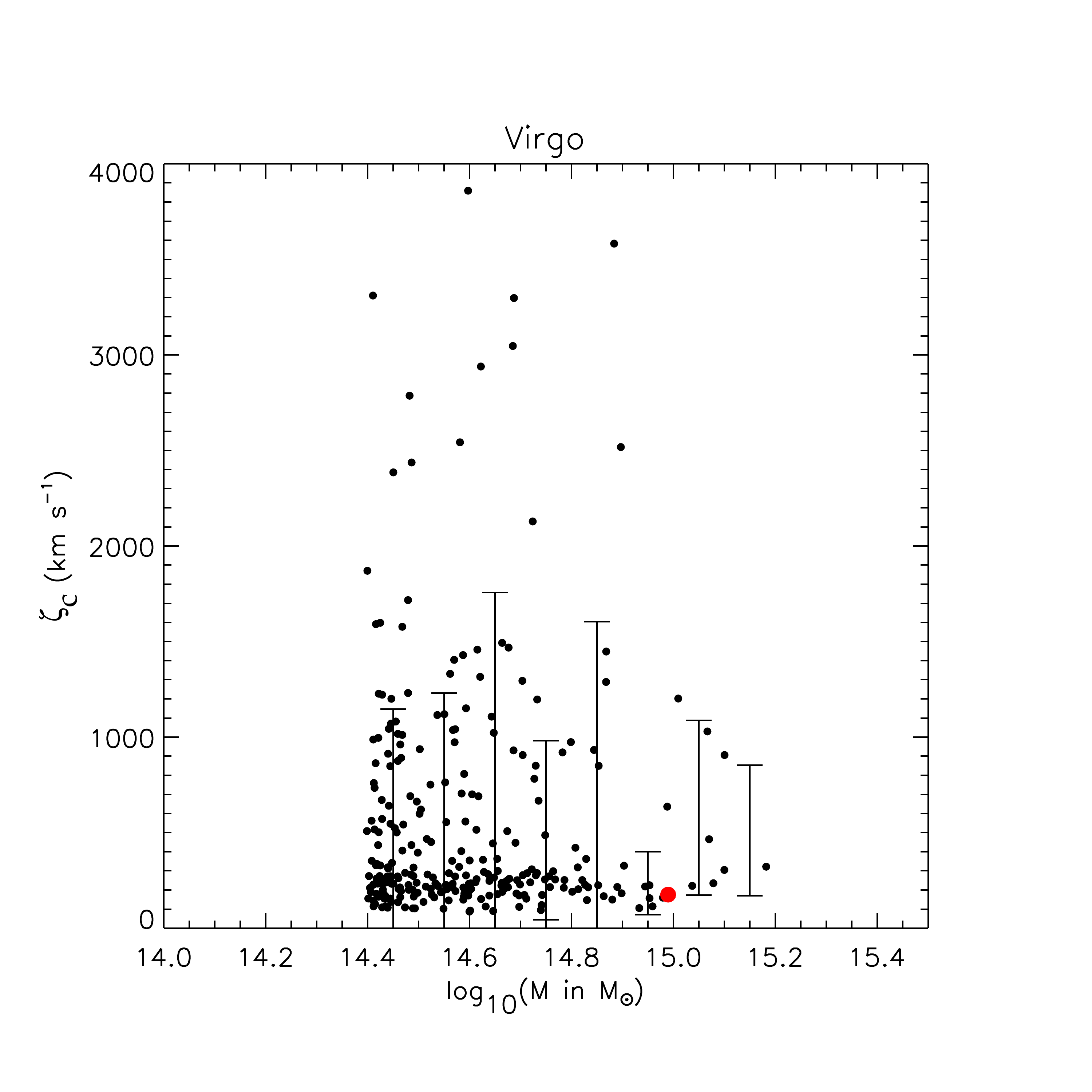}
\hspace{-0.5cm}\includegraphics[width=0.5\textwidth]{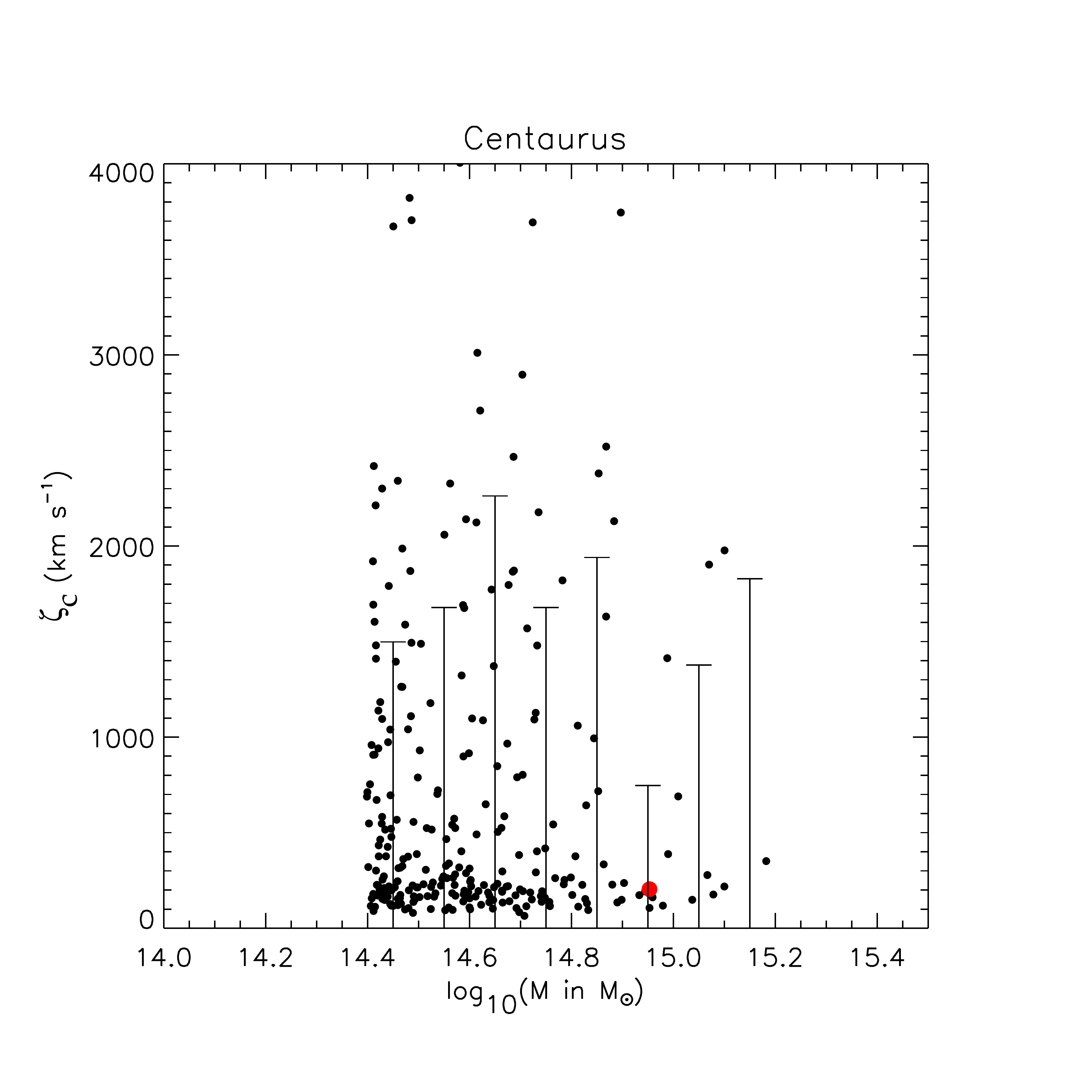}\\
\vspace{-1.2cm}

\hspace{-0.5cm}\includegraphics[width=0.5\textwidth]{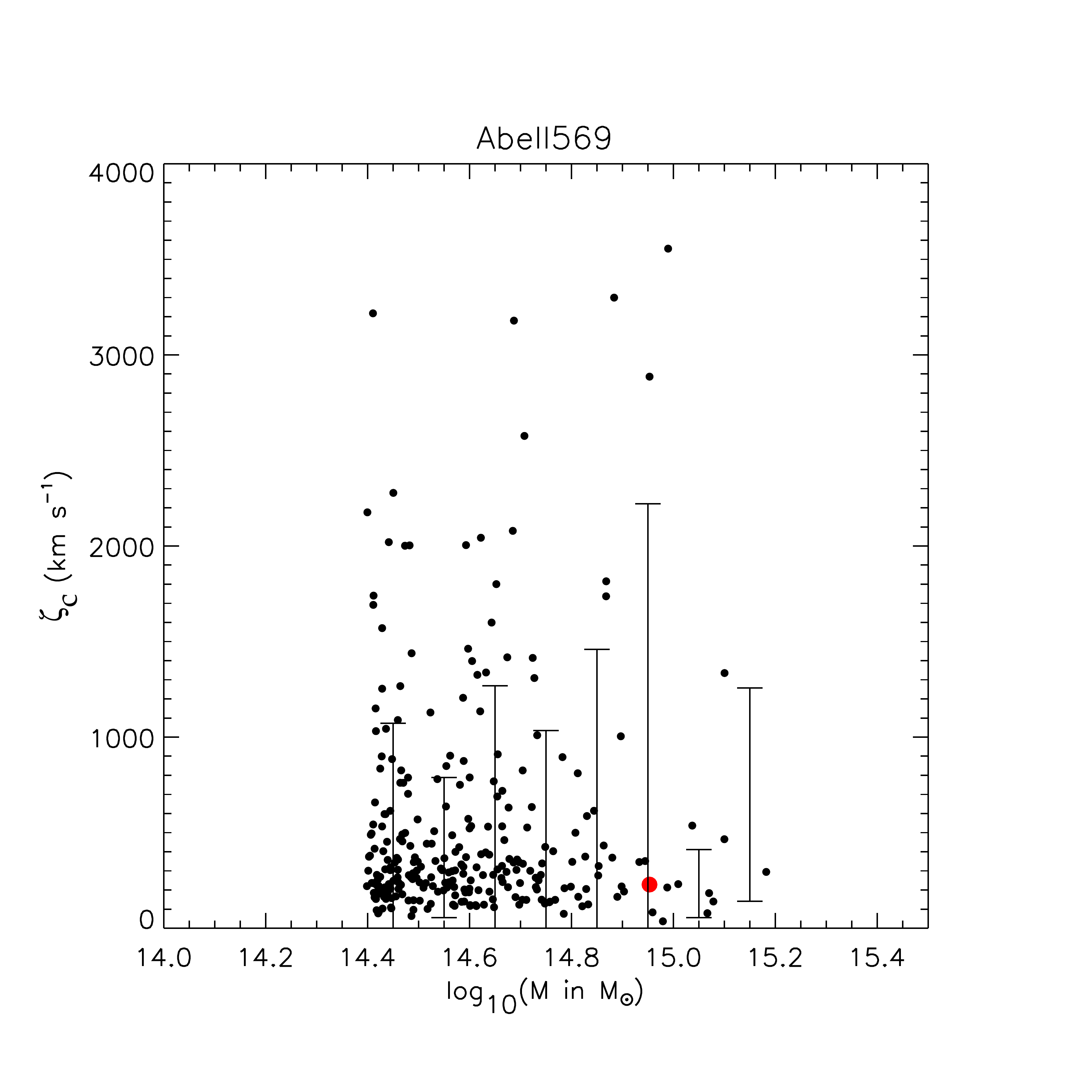}
\hspace{-0.5cm}\includegraphics[width=0.5\textwidth]{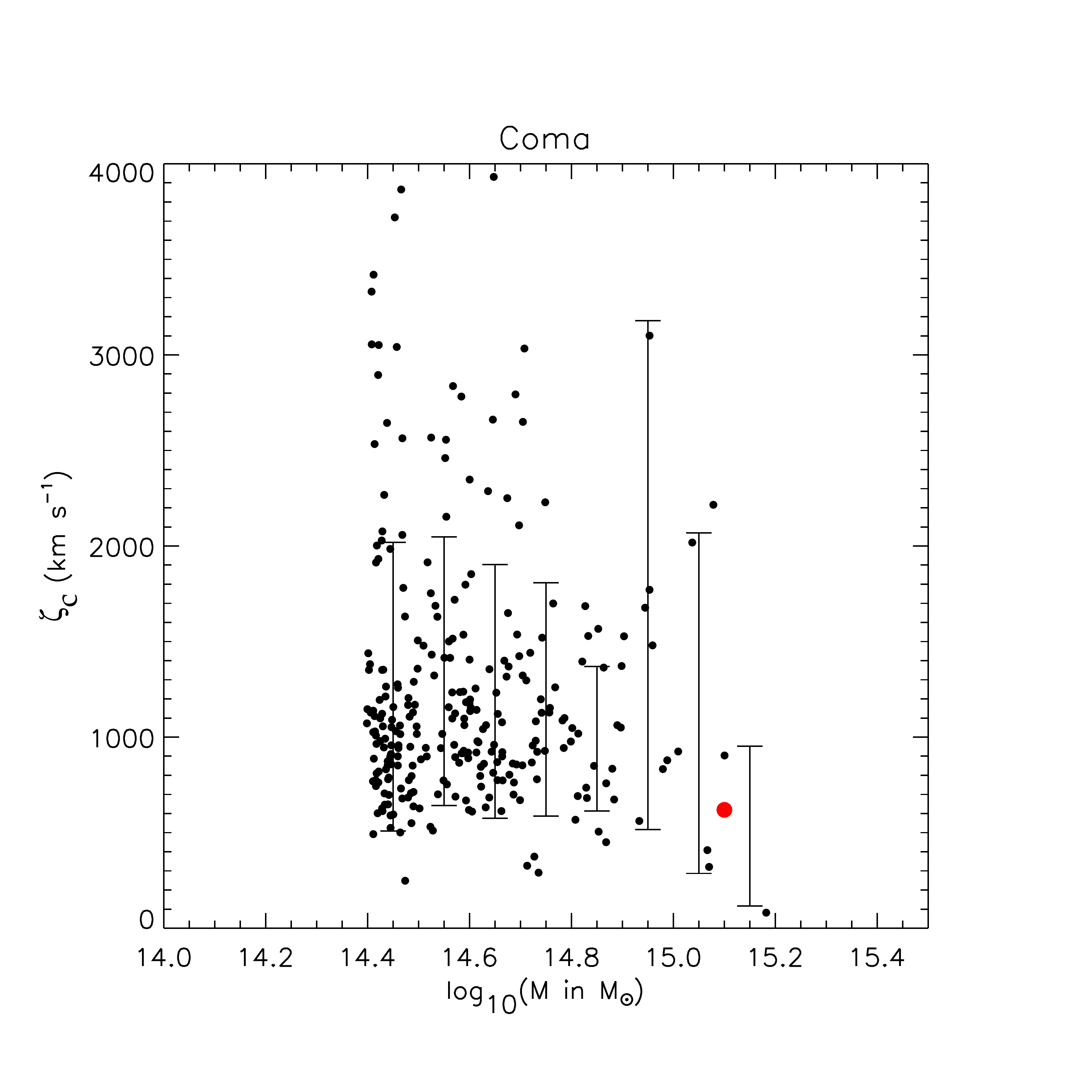}
\vspace{-0.5cm}

\caption{Centered $\zeta$-metric, or $\zeta_c$-metric, values in km~s$^{-1}$ as a function of the halo mass. It measures the difference between the simulated and observed lines-of-sight that include the halos/clusters. Halos that are replicas of the observed clusters, named at the top of each panel, serve as a reference for the shift in the distance of the other halos. The filled red circles give the $\zeta$-metric value obtained for the replicas. The higher $\zeta_c$ is the more different the lines-of-sight are.}
\label{fig:zetac}
\end{figure*}

{$\zeta_c$ values are derived by comparing the observed and simulated datapoints. In Figure \ref{fig:zetac}, these values, their means and standard deviations are represented as filled black circles and error bars per bin of mass for the four clusters. First, the $\zeta$ values for the replicas, given by filled red circles, are on average on the low side compared to $\zeta_c$ values. Second, the scatter and mean of the $\zeta_c$ values vary with the bin and with the compared observed - simulated lines-of-sights.  The smallest mean and scatter are obtained for masses similar to the replica: the filled red circles are at best in the bin of mass with the smallest mean or scatter of $\zeta_c$ values, and at worst in that just next to it. Using twice the current mass bin size gives comparable results but larger uncertainties on mass estimates. Using half the current mass bin size results also in similar values but tends to favor the bin just above that including the $\zeta$ value. We, thus, stick to our current bin size. The $\zeta_c$ values suggest that they might collectively be good proxies for mass estimates. Providing enough simulated halos, deriving the mean and scatter of $\zeta_c$ values per bin of mass associated to these halos and an observational dataset, permits an estimation of the mass of the cluster responsible for the wave in the observational data. It comforts the idea that machine learning techniques could be used to learn this relation between wave and mass to provide cluster mass estimates. This will be investigated in future work.  }

{Table \ref{tbl:mass} gathers the observational virial mass estimates from \citet{2015AJ....149...54T} for the four clusters, as well as those of their replicas in the CLONE. The last column gives the mass bin that corresponds to the $\zeta_c$ values with the smallest mean and scatter. Mass estimates are in good agreement except for Abell 569. The observational mass estimate is one order of magnitude below the $\zeta_c$-metric estimate and the mass of the simulated replica. However, Abell 569 consists of two separated concentrations separated by $\sim$1.5~Mpc \citep{1984A&A...136..178F}. Given the resolution at which CLONE initial conditions are constrained, two concentrations might be replaced with one larger halo. Moreover, the wave signals of two close-by simulated and observed clusters can be entangled affecting their mass estimates. It  highlights a limit of the $\zeta_c$-metric. Clusters need to be sufficiently far apart for their waves to be distinguishable. }

\begin{table}
\centering
\begin{tabular}{lccc}
\hline
Cluster & Observation & CLONE-Replica & $\zeta_c$-metric \\
&\multicolumn{3}{c}{in 10$^{14}$ M$_\odot$}  \\
\hline
Virgo     & 7.01& 9.8 & [7.9-10] \\  
Centaurus & 10.8 & 9.0 &[7.9-10] \\
Abell 569 & 0.67 &  9.0 & [10-13] \\
Coma      & 15.9  & 12.6  & [13-16]   \\
\hline
\end{tabular}
\caption{Cluster mass estimates from observations \citep{2015AJ....149...54T}, from the replica and using the $\zeta_c$-metric to compare observed and {several} simulated lines-of-sight.}
\label{tbl:mass}
\end{table}


\section{Conclusions}

Galaxy clusters are good cosmological probes provided their mass estimates are accurately determined. Fueled with large imaging surveys, stacked weak lensing is the most promising mass estimate method though it provides estimates within relatively small radii. Given the large amount of accompanying redshift and spectroscopic data overlapping the imaging surveys, we must take the opportunity to calibrate also with a reasonable accuracy a method based on galaxy dynamics. Two independent measures hold indeed better constraints on the cosmological model. The infall zones of galaxy clusters are probably the less sensitive to baryonic physics, thus mostly shielded from challenging systematics, and probe large radii. These manifestations of a tug of war between gravity and dark energy provide a unique avenue to test modified gravity theories when comparing resulting mass estimates to those from stacked weak lensing measurements. Combined with stacked weak lensing results, they might even yield evidence that departure from general relativity on cosmological scales is responsible for the expansion acceleration.\\

The accurate calibration of the relation between infall zones properties and cluster masses starts with careful comparisons between cosmological simulations and observations. In this paper, we thus present our largest and highest resolution Constrained Local \& Nesting Environment Simulation (CLONE) built so far to reproduce numerically our cosmic environment. This simulation stems from initial conditions constrained by thepeculiar velocities of local galaxies. By introducing this cosmological dark matter CLONE of the local large-scale structure with a particle mass of $\sim$10$^9$M$_\odot$ within a $\sim$738~Mpc box, we have sufficient resolution to study the effect of the gravitational potential of massive local halos onto the velocity of (sub)halos. We can also compare with that of their observational cluster counterparts. \\

Velocity waves stand out in radial peculiar velocity - distance to a box-centered synthetic observer diagram. Lines-of-sight that include velocity waves, caused by the most massive dark matter halos of the CLONE simulation and those born from their observed local cluster counterparts are in agreement, especially the clusters the closest to us that are the best constrained (e.g. Virgo, Centaurus). Secondary waves due to smaller groups in (quasi) the same line-of-sight as the most massive clusters stand out equally even though they are further into the non-linear regime. Indeed, prior to full non-linear evolution to the z=0 state, assuming $\Lambda$CDM, CLONE initial conditions are constrained with solely the linear theory, a power spectrum and highly uncertain and sparse local peculiar velocities. The visual matching between the simulated and observed lines-of-sight is confirmed with 2D-Kolmogorov Smirnov (KS) statistic values and tests as well as with our own $\zeta$-metric. Contrary to the 2D-KS statistic, the $\zeta$-metric takes into account the real distance of galaxies along the entire lines-of-sight (not only the studied fractions). The $\zeta$-metric is however more sensitive to the fact that observational uncertainties are not taken into account in these metrics. The two metrics appear to be complementary. They show that the closest clusters have the best reproduced lines-of-sight. The lines-of-sight of clusters at the edges of the constrained region and even slightly beyond are reproduced by the simulation to a smaller extent (lower mass limit reached).  \\

Additionally, a Gaussian-plus-continuum fit to the envelope of the positive-half of all the velocity waves born from dark matter halos more massive than 2~10$^{14}$M$_\odot$ in the simulation reveals both the variety and complexity of the potential wells as well as the correlation of the fit parameters with the halo masses. Overall, the Gaussian amplitude is mostly linked to the halo mass, but for a few exceptions, with a residual scatter. Although the Gaussian standard deviation is not always correlated with the mass, it can be slightly correlated with the Gaussian amplitude thus with the mass. The continuum is certainly an interesting parameter to consider as it permits splitting the halos into different classes. Each continuum value seems to drive a given correlation between the Gaussian amplitude and the halo mass and, to a smaller extent, with the Gaussian standard deviation. To summarize, parameter fits are completely interdependent, a global fit to the velocity wave is then the best approach to obtain a first rough cluster mass estimate.\\

We further derive centered $\zeta$-metric, or $\zeta_c$-metric values for observed and simulated lines-of-sight including different mass galaxy clusters. Simulated halos are shifted in distance to match that of the observed clusters, hence the term ``centered'' $\zeta$-metric.  The $\zeta_c$ values per bin of mass present a minimum mean and variance in agreement with the $\zeta$ value obtained when comparing simulated and observed lines-of-sight of the cluster and its replica. {This suggests that the $\zeta_c$-metric can provide the mass range estimate of a cluster given an observational dataset that includes its wave and several simulated cluster re-centered lines-of-sight}. It also comforts the idea that machine learning techniques should be able to learn wave-type and associated masses to finally give cluster mass estimates. We will investigate this in future work.\\

First and foremost, this work confirms the potential of the velocity wave technique to get massive cluster mass estimates and  test gravity and cosmological models. Our CLONES, with the first shown reproduction of observed lines-of-sight including velocity waves, could in the near future provide the zero point of galaxy infall kinematic technique calibrations \citep{2013MNRAS.431.3319Z}. A Bayesian inference model embedding a machine learning technique built and trained on random simulated galaxy surveys that is then applied to both constrained simulated and observed galaxy surveys must recover the same local velocity waves and corresponding mass estimates to be validated. Our CLONES will moreover allow minimizing observational biases as any real environmental and cluster property will be reproduced for one-to-one comparisons. Local kinematic mass estimates can then become accurate. Once compared with other techniques of local galaxy cluster mass estimates, they will permit the calibration of the zero-point of these other techniques to be applied to further-and-further away clusters. \\


\section*{Acknowledgements}
{The authors would like to thank the referee for their comments that help improve the quality of this manuscript}. The authors acknowledge the Gauss Centre for Supercomputing e.V. (www.gauss-centre.eu) and GENCI (https://www.genci.fr/) for funding this project by providing computing time on the GCS Supercomputer SuperMUC-NG at Leibniz Supercomputing Centre (www.lrz.de) and Joliot-Curie at TGCC (http://www-hpc.cea.fr), grants ID:~22307/22736 and A0080411510 respectively. This work was supported by the grant agreements ANR-21-CE31-0019 / 490702358  from the French Agence Nationale de la Recherche / DFG for the LOCALIZATION project and ERC-2015-AdG 695561 from the European Research Council (ERC) under the European Union's Horizon 2020 research and innovation program for the ByoPiC project (https://byopic.eu). This work was supported by the Programme National Cosmologie et Galaxies (PNCG) of CNRS/INSU with INP and IN2P3, co-funded by CEA and CNES. KD acknowledges support by the COMPLEX project from the ERC under the European Union's Horizon 2020 research and innovation program grant agreement ERC-2019-AdG 882679. The authors thank the referee for their comments. JS thanks Marian Douspis for useful comments, the ByoPiC team, the IAP `Origin and evolution of galaxies' team and her CLUES collaborators for  discussions. NM acknowledges funding by the European Union through a Marie Sk{\l}odowska-Curie Action Postdoctoral Fellowship (Grant Agreement: 101061448, project: MEMORY).

\bibliographystyle{aa}

\bibliography{biblicompletenew}

\newpage
\begin{appendix}
\section{Comparisons with SIBELIUS-DARK}

\begin{figure}
\center
\vspace{-1.cm}

\includegraphics[width=0.45\textwidth, trim={1cm 0 1cm 0},clip]{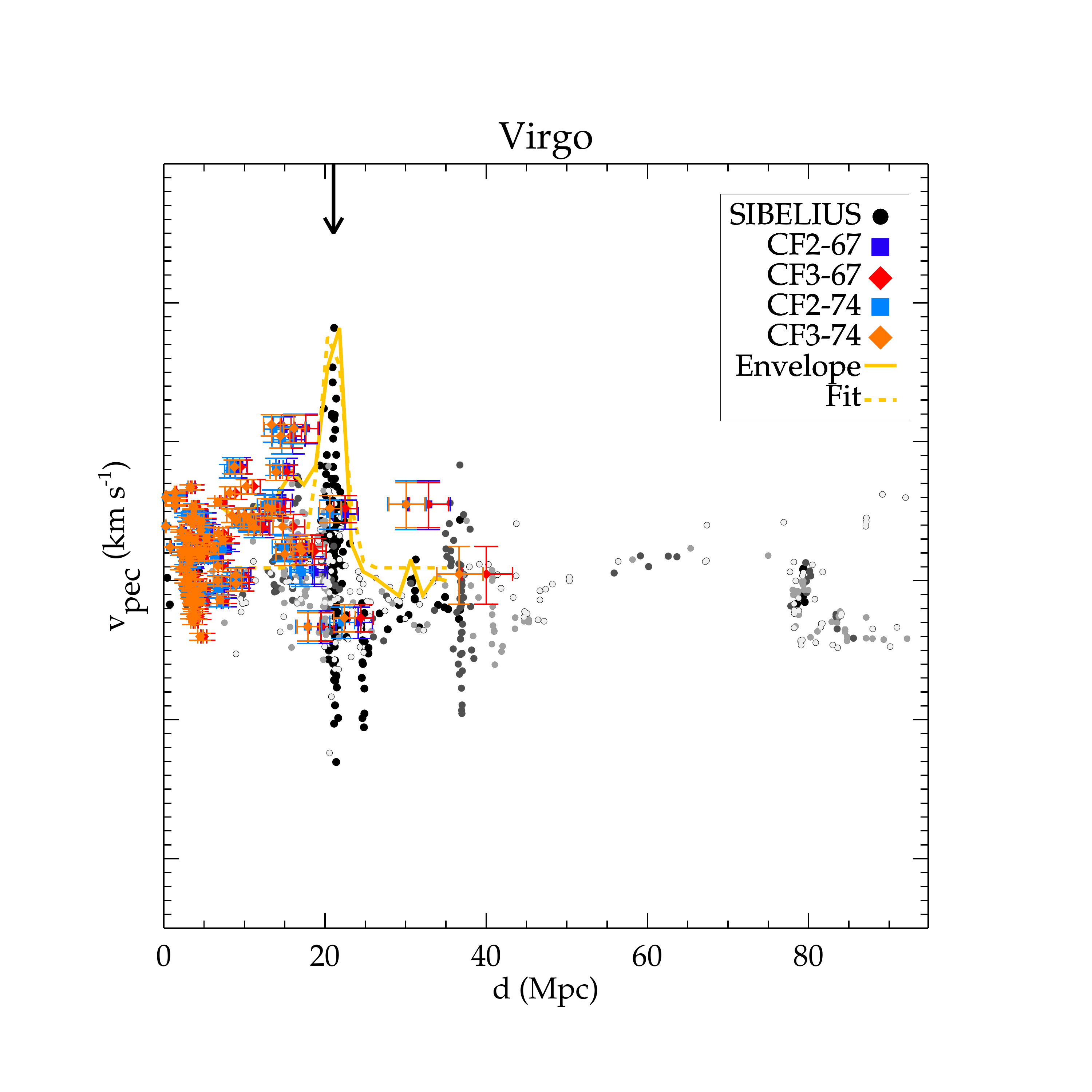}\vspace{-1.5cm}

\includegraphics[width=0.45\textwidth, trim={1cm 0 1cm 0},clip]{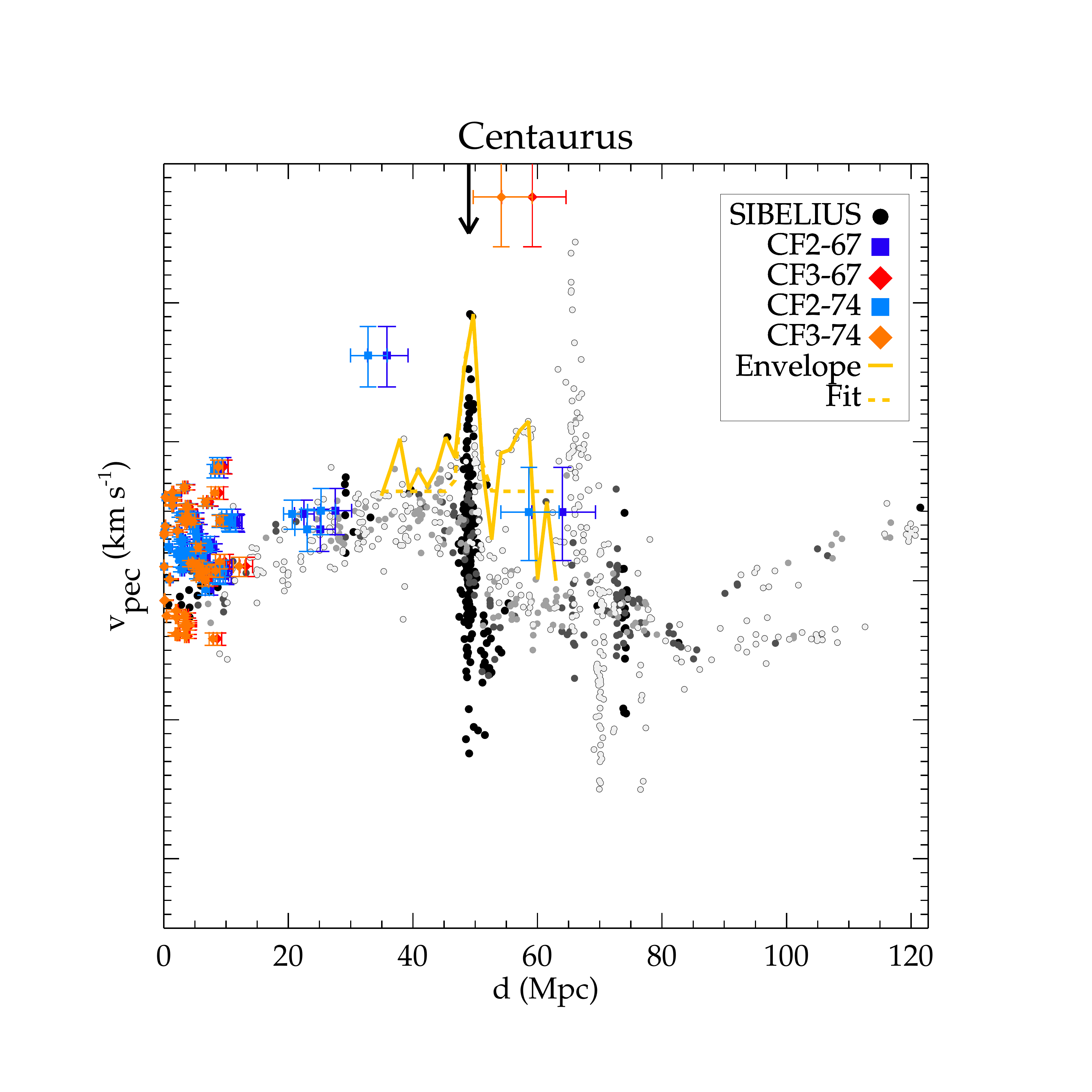}\vspace{-1.5cm}

\includegraphics[width=0.45\textwidth, trim={1cm 0 1cm 0},clip]{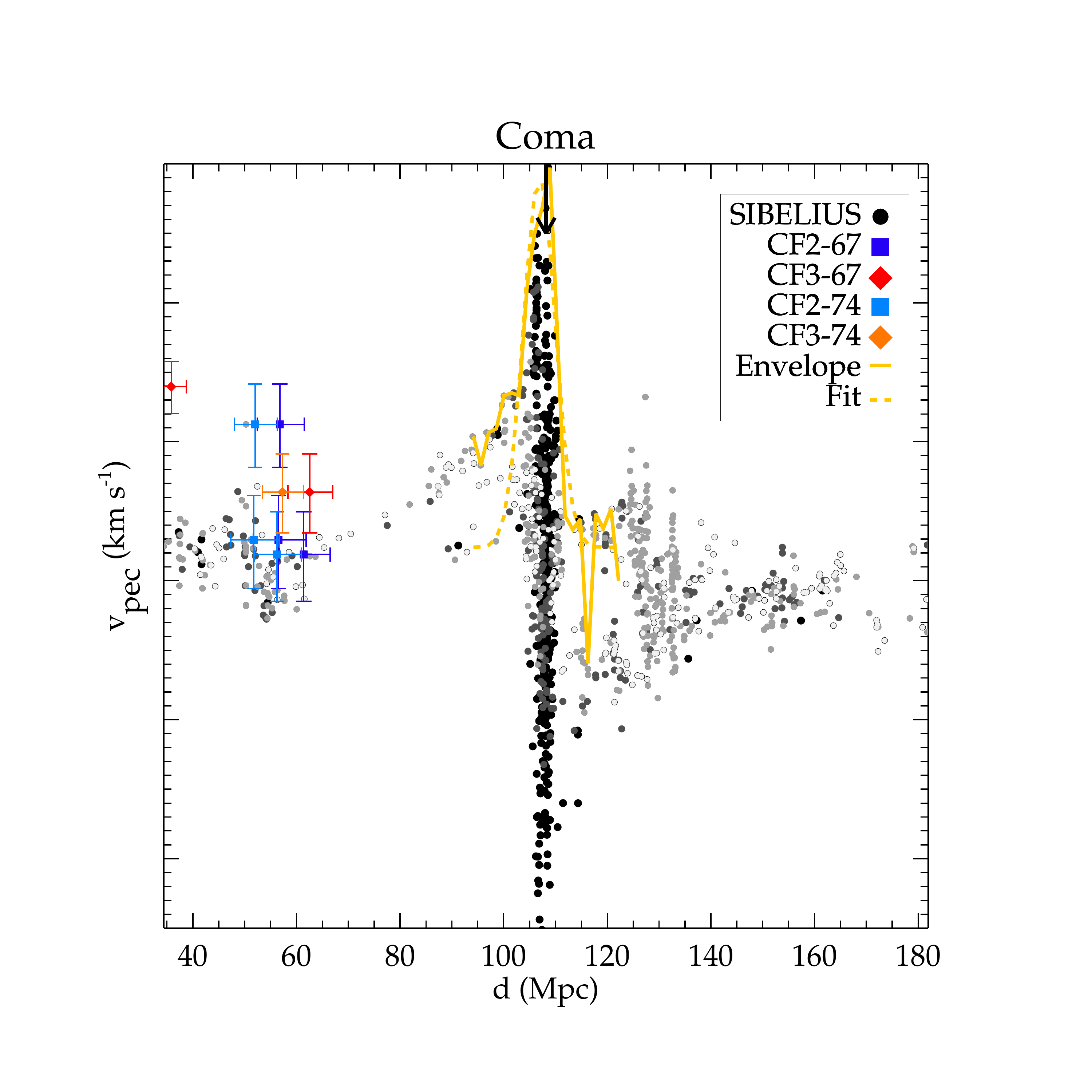}
\vspace{-0.5cm}

\caption{Same as Figure \ref{fig:waves} but for the SIBELIUS-DARK clusters \citep{2022MNRAS.512.5823M}.}
\label{fig:sibelius}
\end{figure}

{This appendix compares the observed lines-of-sight to those simulated in the SIBELIUS-DARK constrained simulations \citep{2022MNRAS.512.5823M}. Contrary to CLONES, this simulation is constrained with a complete, to a given limit, observational redshift survey. Given the halos identified in \citet{2022MNRAS.512.5823M} and the observational dataset with the required uncertainty threshold available to compare them to, three clusters are selected: Virgo, Centaurus and Coma. The SIBELIUS-DARK and observed lines-of-sight are shown in Figure \ref{fig:sibelius}. Lines-of-sight are in good agreement with the observational datapoints except for Coma-SIBELIUS. However, a smaller group in front of Coma-SIBELIUS matches some observational datapoints. Because of this absence of match with the main cluster wave, Coma is removed from further comparisons.}

{Tables \ref{tbl:KsStatMetricsibelius} gather the KS-statistic and $\zeta$-metric values. The two cluster lines-of-sight seem to be reproduced in a comparable way in CLONE and SIBELIUS with a tiny advantage for Virgo-CLONE (Virgo-SIBELIUS) when using the KS-statistic ($\zeta$-metric).  However, SIBELIUS-DARK line-of-sight has twice more simulated datapoints to compare to a comparable number or less observational datapoints than for CLONE line-of-sight. This highlights a limitation of the $\zeta$-metric, which favors simulated lines-of-sight with a larger number of simulated datapoints to match with a smaller number of observed datapoints. There are indeed more choices. This results in a higher probability of finding a smaller minimum.This difference is partly due to the semi-analytical model used in SIBELIUS-DARK. Although we restricted this sample in mass to match our dark matter subhalo list, samples are not strictly identical. In addition, replicas are not located exactly at the same positions in CLONE versus SIBELIUS-DARK, implying slightly different matched lines-of-sight with observations. Synthetic line-of-sight replicas are not shifted to match the observed lines-of-sight but simply overplotted. }

{In Table \ref{tbl:masssibelius}, the $\zeta_c$-metric provides an estimate of the mass of the clusters given the lines-of-sight matching SIBELIUS-DARK ones. While for Virgo the mass range is the same as for CLONE, it is higher for Centaurus. The mix between two waves in the Centaurus-SIBELIUS line-of-sight might be the cause. The $\zeta_c$-metric is potentially biased high when two halo signatures are entangled, as shown for Abell 569-CLONE.} \\

\begin{table}
\centering
\begin{tabular}{l@{ }@{ }ccccc}
\hline
Cluster &\multicolumn{2}{c}{SIBELIUS/CF2}  &  \multicolumn{2}{c}{SIBELIUS/CF3} & Random/CF3 \\
Cylinder  & 10 & 2.5 & 10  & 2.5 & 2.5 \\
radius& Mpc & Mpc & Mpc & Mpc & Mpc\\
\hline
Virgo     & 0.80 & 0.88 & 0.77 & 0.87  &/\\  
Centaurus & 0.85 & 0.90 & 0.91 & 0.88  & /\\
\hline
\hline
Virgo     &  129 &  208  & 116 & 207  & 11200 \\  
Centaurus & 171  & 312 & 181  & 279  & 11365
\end{tabular}
\caption{Kolmogorov-Smirnov statistic (first two lines) and $\zeta$-metric in km~s$^{-1}$ (last two lines) like in Table \ref{tbl:KsStat} and Table \ref{tbl:MetricStat} but for SIBELIUS-DARK clusters.}
\label{tbl:KsStatMetricsibelius}
\end{table}

\begin{table}
\centering
\begin{tabular}{lccc}
\hline
Cluster & Observation & SIBELIUS-Replica & $\zeta_c$-metric \\
&\multicolumn{3}{c}{in 10$^{14}$ M$_\odot$}  \\
\hline
Virgo     & 7.01& 3.5 & [7.9-10] \\  
Centaurus & 10.8 & 4.5 &[13-16] \\
\hline
\end{tabular}
\caption{Cluster mass estimates from observations \citep{2015AJ....149...54T}, the SIBELIUS-replica and the $\zeta_c$-metric.SIBELIUS-DARK mass estimates are M$_{200}$. They are $\sim$1.3 times smaller than M$_{vir}$  \citep{2016MNRAS.460.2015S} .}
\label{tbl:masssibelius}
\end{table}

\end{appendix}

 \label{lastpage}
\end{document}